\documentstyle[preprint,aps,eqsecnum]{revtex}
\tighten
\def\a{\alpha}
\def\b{\beta}
\def\g{\gamma}

\def\l{\lambda}
\def\s{\sigma}

\def\Th{\Theta}
\def\o{\over}
\newcommand{\tr}{\mbox{tr}}
\begin{document}
\preprint{LPTHE-Paris-95-14/DEMIRM-Paris-95009/hep-th/9504098}

\bigskip

\title{\Large{\bf  STRING THEORY IN \\
  COSMOLOGICAL  SPACETIMES}}
\author{{\bf H.J. de Vega$^{(a)}$
 and N. S\'anchez$^{(b)}$}\\}

\bigskip

\address
{  (a)  Laboratoire de Physique Th\'eorique et Hautes Energies,
Universit\'e Pierre et Marie Curie (Paris VI) et Universit\'e Denis
Diderot (Paris VII),
Tour 16, 1er. \'etage, 4, Place Jussieu
75252 Paris, cedex 05, {\bf France}.  Laboratoire Associ\'{e} au CNRS URA280.\\
 (b) Observatoire de Paris, DEMIRM, 61, Avenue de l'Observatoire,
75014 Paris, {\bf France}.  Laboratoire Associ\'{e} au CNRS URA336,
Observatoire de Paris et \'{E}cole Normale Sup\'{e}rieure.\\}

\bigskip

\date{April 1995}

\maketitle

\bigskip

\bigskip

\bigskip

\begin{center}

Lectures delivered at the Erice School

``CURRENT TOPICS IN ASTROFUNDAMENTAL PHYSICS''

4-16 September 1994, to appear in the Proceedings edited by  N. S\'anchez.

\end{center}
\newpage

\title{\bf  STRING THEORY IN \\
 COSMOLOGICAL  SPACETIMES}
\author{{\bf H.J. de Vega$^{(a)}$
 and N. S\'anchez$^{(b)}$}}
\address
{  (a)  Laboratoire de Physique Th\'eorique et Hautes Energies,
Universit\'e Pierre et Marie Curie (Paris VI) et Universit\'e Denis
Diderot (Paris VII),
Tour 16, 1er. \'etage, 4, Place Jussieu
75252 Paris, cedex 05, France.  Laboratoire Associ\'{e} au CNRS URA280.\\
 (b) Observatoire de Paris, DEMIRM, 61, Avenue de l'Observatoire,
75014 Paris, France.  Laboratoire Associ\'{e} au CNRS URA336,
Observatoire de Paris et \'{E}cole Normale Sup\'{e}rieure.}
\maketitle
\begin{abstract}
Progress on string theory in curved spacetimes since 1992 are
reviewed. After a short introduction on strings in Minkowski and curved
spacetimes, we focus on strings in cosmological spacetimes.

The classical behaviour of strings in FRW and inflationary
spacetimes is now understood in a large extent from
 various types of explicit  string solutions.
Three different types of behaviour appear in cosmological spacetimes:
{\bf unstable, dual} to unstable and {\bf stable}.
For the unstable strings, the energy and size grow  for large    scale
factors $R \to \infty$,  proportional to $R$. For the dual to unstable
strings, the energy and size blow up for $R\to 0$ as $1/R$. For stable
strings,  the energy and proper size are bounded. (In Minkowski
spacetime, all string solutions are of the stable type).

Recent progress on self-consistent solutions to the Einstein equations
for string dominated universes is reviewed. The energy-momentum
tensor for a gas of strings is then considered  as source of the
spacetime geometry and from the above string behaviours  the string
equation of state is determined. The  self-consistent
string  solution  exhibits the realistic matter dominated behaviour
$ R \sim (X^0)^{2/3}\; $ for large times and the radiation dominated
 behaviour $ R \sim (X^0)^{1/2}\; $ for early times.

Finally, we report on the exact
integrability of the string  equations plus the constraints in de
Sitter spacetime  that allows to  systematically find {\bf
exact} string solutions  by soliton methods  and the multistring solutions.
 {\bf Multistring solutions} are a new  feature  in curved spacetimes.
  That is,  a single world-sheet  simultaneously describes
many   different and independent strings. This phenomenon has no
analogue in flat spacetime and  appears as a
consequence of the coupling of the strings with the spacetime geometry.
\end{abstract}

\newpage

\section{\bf Introduction}

Since the previous Erice School on `String Quantum Gravity' a host of
impressive developments has taken place on strings in curved
spacetime. For the state of the art in 1992 we refer to the 1992
Proceedings \cite{eri92} where the programme on string quantization in
curved spacetimes initiated by us in 1987 \cite{dvs87} is reviewed.

A consistent  quantum theory of gravity is the strongest motivation
for string theory and hence to study strings in curved
spacetime\cite{eri92}. As stressed in the sec. 1 of the 1992 lectures,
a quantum theory of gravity must be a theory able to
describe all physics below the Planck scale  $M_{Planck}  =  \hbar c /
G  =  1.22 \,  10^{16} $Tev.  That means that a sensible  theory of
quantum gravity is necessarily part of a unified theory of all interactions.
 {\bf Pure gravity} (a model   containing only gravitons) cannot be a
physical and realistic quantum theory.
To give an example,
  a theoretical prediction for graviton-graviton scattering at energies of
 the order of $M_{Planck}$   must include all particles produced in a real
 experiment. That is, in practice, all existing particles in nature, since
 gravity couples to all matter.

String theory is therefore a serious candidate for a quantum
description of gravity since it provides a unified model of all
interactions overcoming at the same time the nonrenormalizable
character of quantum fields theories of gravity.

\bigskip

The present lectures deal mainly  with strings in cosmological
spacetimes;  substantial results were achieved in this field since
1992. The classical behaviour of strings in FRW and inflationary
spacetimes is now understood in a large extent\cite{cos}. This understanding
followed the finding of various types of exact and numerical string
solutions in  FRW and
inflationary spacetimes\cite{dms} -\cite{din}. For inflationary
spacetimes, the exact
integrability of the string propagation equations plus the string
constraints in de Sitter spacetime \cite{prd} is indeed an
important help.
This allowed to  systematically find {\bf exact} string solutions  by soliton
methods using the linear system associated to the problem
(the so-called dressing method in soliton theory)
and the multistring solutions \cite{dms} -\cite{igor}.

  In summary, three different types of
behaviour are exhibited by the string solutions in cosmological
spacetimes: {\bf unstable, dual} to unstable and {\bf stable}.
For the unstable strings, the energy and size grow  for large    scale
factors $R \to \infty$,  proportional to $R$. For the dual to unstable
strings, the energy and size blow up for $R\to 0$ as $1/R$. For stable
strings, the energy and proper size are bounded. (In Minkowski
spacetime, all string solutions are of the stable type).  The equation
of state for these string behaviours take the form
\begin{itemize}
\item (i) {\bf unstable}  for $ R \to \infty
 \;   p_u =  -E_u/(d-1) < 0 $
\item (ii) {\bf dual to unstable}  for $ R \to 0
 , \; p_d = E_d/(d-1) > 0  $ .
\item (iii) {\bf stable} for $ R \to \infty ,
 \;  p_s = 0 $ .
\end{itemize}
Here $E_u$ and  $E_d$ stand for the corresponding string energies and
$d-1$  for the number of spatial dimensions where the string
solutions lives. For example, $d-1 = 1$ for a straight string,
 $d-1 = 2$ for a ring string, etc. This number $d$ is obviously less or equal
than the number of  spacetime dimensions $D$.

As we see above,
the dual to unstable string behavior leads to the same equation of
state than radiation   (massless particles or hot matter). The stable
string behavior leads to the  equation of
state of massive particles (cold matter). The unstable string behavior
is a purely `stringy' phenomenon. The fact that in entails a negative
pressure is however  physically acceptable.
For a gas of strings, the unstable string behaviour dominates in
inflationary universes when  $ R \to \infty$ and the   dual to
unstable string behavior dominates for  $ R \to 0 $.

The unstable strings
correspond to the critical case of the so called {\it coasting universe}
\cite{ell,tur}. That is, classical strings provide a {\it concrete}
realization of such cosmological models.
The `unstable' behaviour is called  `string stretching'
in the cosmic string literature \cite{twbk,vil}.

It must be stressed that while  time evolves, a {\bf given} string
solution may exhibit two and even three of the above regimes one after
the other (see sec. III).
 Intermediate behaviours are also observed in ring solutions
\cite{dls,din}. That is,
$$
P = (\g - 1)\; E \quad {\rm with~} -{1 \o {d-1}}< \g - 1< +{1 \o {d-1}}
$$

Another new  feature appeared in curved spacetimes: {\bf multistring
solutions}.  That is  a single world-sheet  simultaneously describes
many   different and independent strings. This phenomenon has no
analogue in flat spacetime.
This is a new feature appearing as a
consequence of the interaction of the strings with the spacetime geometry.

The world-sheet time $\tau$ turns out to be an multi-valued
function of the target string time $X^0$ (which can be
the cosmic time $T$, the conformal time $\eta$ or for de Sitter
universes it can be the hyperboloid time $q^0$).
Each branch of $\tau$ as a
function of $X^0$ corresponds to a different string. In flat spacetime,
multiple string solutions are necessarily described by multiple
world-sheets. Here, a single world-sheet describes  one string,
several strings or even an infinite number of
different and independent strings as a consequence of the coupling with the
spacetime geometry. These strings do not interact among themselves; all the
interaction is with the curved spacetime.  One can decide  to study
separately each of them (they are all different)
or consider all the infinite strings together.

Of course, from our multistring solution, one {\it could} just
choose only one interval in $ \tau $ (or a subset of intervals  in $
\tau $) and describe just one string (or several). This will be just a
{\bf truncation} of the solution.

The really remarkably fact
is that all these infinitely many strings come {\bf naturally together}
when solving the string equations in de Sitter spacetime as we did in
refs. \cite{dms} -\cite{dls}.

Here, interaction among the strings (like splitting and merging) is neglected,
the only interaction is with the curved background.

The study of string propagation in curved spacetimes provide essential
clues about the physics in this context but is clearly not the end of
the story. The next step  beyond the investigation of {\bf test}
strings, consist in finding {\bf self-consistently} the geometry from
the strings as matter sources for the Einstein equations
or better the string effective equations (beta functions).
This goal is achieved in ref.\cite{cos} for cosmological spacetimes at the
classical level. Namely, we used  the energy-momentum tensor for a gas
of strings as source for the Einstein equations and we solved them
self-consistently.

To write the string equation of state we used the behaviour of the
string solutions in cosmological spacetimes.
Strings continuously evolve from one type of behaviour to another, as
is explicitly shown by our solutions \cite{prd} -\cite{dls}. For
intermediate values of $ R $, the  equation of state for gas of free strings
 is clearly complicated but a formula of the type:
\begin{equation}
\rho = \left( u_R \; R + {{d} \over R} + s \right) {1 \over
{R^{D-1}}} \label{rogenI}
\end{equation}
 where
\begin{eqnarray}
\lim_{R\to\infty} u_R = \cases{ 0 \quad & {\rm FRW } \cr
  u_{\infty} \neq 0 & {\rm Inflationary } \cr}
\end{eqnarray}
 This equation of state is qualitatively
correct for all $ R $ and becomes exact for $ R \to 0 $ and $ R \to
\infty $ . The parameters
$u_R , d$ and $ s $ are positive constants and the $u_R$
varies smoothly with $R$.

The pressure associated to the energy density (\ref{rogenI}) takes then
the form
\begin{equation}
p  = {1 \over {D-1}} \left( {d \over R} -  u_R \; R\right) {1 \over
{R^{D-1}}} \label{pgenI}
\end{equation}

Inserting  this source into the Einstein-Friedman equations leads to a
self-consistent solution  for  string dominated universes (see sec. IV)
\cite{cos}.
This solution exhibits the realistic matter dominated behaviour
$ R \sim (X^0)^{2/(D-1)}\; $ for large times and the radiation dominated
 behaviour $ R \sim (X^0)^{2/D}\; $ for early times.

For the sake of completeness we analyze in sec. IV
 the effective string equations \cite{cos}.
These equations have been extensively treated in the
literature \cite{eqef} and they are not our central aim.

It must be noticed that there is no satisfactory
derivation of inflation in the context of the effective string equations.
 This does not mean that string
theory is not compatible with inflation, but that the effective string
action approach {\it is not enough} to describe inflation. The
effective string equations are a low energy field theory approximation
to string theory containing only the {\it massless} string modes.
The vacuum energy scales to start inflation are typically of the order
of the Planck mass where the effective string action approximation
breaks down. One must also consider the {\it massive} string modes (which
are absent from  the effective string action) in order to properly get
the cosmological condensate yielding inflation.
De Sitter inflation does not emerge as a solution of the
the effective string equations.

In conclusion, the effective
string action (whatever be the dilaton, its potential and the
central charge term) is not the appropriate framework in which to
address the question of string driven inflation.

Early cosmology (at the Planck time) is probably the best place to test
string theory. In one hand the quantum treatment of gravity is
unavoidable at such scales
and in the other hand, observable cosmological consequences are
derivable from the inflationary stage.
The natural gravitational background is  an
inflationary universe as de Sitter spacetime. Such geometries are not
string vacua. This means that conformal and Weyl symmetries are broken at
the quantum level. In order to quantize  consistently  strings in such
case, one must enlarge the physical phase space including, in
particular, the  factor  $\exp\phi(\s,\tau)$
in the world-sheet metric [see eq.(\ref{liou})]. This is a very
interesting and open problem. Physically, the origin of such
difficulties in quantum string cosmology comes from the fact that one
is not dealing with an {\bf empty} universe since a cosmological spacetime
 necessarily contains matter. In the other hand, conformal
field theory techniques are only adapted to backgrounds for which the
beta functions are identically zero, i. e. sourceless geometries.

The outline of these  lectures is as follows. Section II
presents an introduction to strings in curved spacetimes including
basic notions on classical and quantum strings in Minkowski spacetime
and introducing the main physical string magnitudes: energy-momentum
and invariant string size.  Section III deals with the
string propagation and the string energy-momentum tensor in
cosmological spacetimes. (In sections III.A, III.B and III.C we treat the
straight strings, ring strings and generic strings
respectively and derive the
corresponding string equations of state). In section IV we treat
self-consistent string cosmology including the string equations
of state. (Section IV.A deals with
general relativity, IV.B with the  string thermodynamics).
Section V discuss  the effective (beta functions) string equations
in the cosmological perspective and the search of inflationary solutions.
Finally, in sec. VI, we briefly review the systematic construction of
string solutions in de Sitter universe {\it via} soliton methods and
the new feature of multistring solutions.

\section{\bf  Introduction. Strings in Curved and Minkowski Spacetimes.}

 Let us consider bosonic strings (open or closed) propagating in a
curved D-dimensional spacetime defined by a metric $G_{AB}(X),
0 \leq A,B \leq D-1$.
  The action can be written as
 \begin{equation}\label{accion}
    S  = {{1}\o{2 \pi \a'}} \int d\s d\tau \sqrt{g}\,  g_{\a\b}(\s,\tau) \;
 G_{AB}(X) \,
 \partial^{\a}X^A(\s,\tau) \, \partial^{\b}X^B(\s,\tau)
 \end{equation}
 Here  $ g_{\a\b}(\s,\tau)$  ( $0 \leq \a, \b \leq 1$ ) is the metric in the
 worldsheet, $\a'$ stands for the string tension. As in flat spacetime, $\a'
 \sim (M_{Planck})^{-2} \sim ( l_{Planck})^2$  fixes the scale in the theory.
  There are no other free parameters like coupling constants in string theory.

We will start considering given gravitational backgrounds  $G_{AB}(X)$.
That is, we start to investigate {\em test} strings propagating on a
given spacetime. In section IV,
the back reaction problem will be studied. That is, how the strings
may act as source of the geometry.

String propagation in massless backgrounds other
than gravitational (dilaton, antisymmetric tensor) can be investigated
analogously.

 The string action (\ref{accion}) classically enjoys Weyl invariance
 on the world sheet
 \begin{equation}\label{weyl}
  g_{\a\b}(\s,\tau) \to \l(\s,\tau) \,  g_{\a\b}(\s,\tau)
 \end{equation}
 plus the reparametrization invariance
 \begin{equation}\label{reparam}
 \s \to  \s'  =  f(\s,\tau) \qquad , \qquad \tau \to \tau'   =  g(\s,\tau)
 \end{equation}
Here $ \l(\s,\tau),  f(\s,\tau)$ and $  g(\s,\tau)$ are arbitrary functions.

         The dynamical variables being here the string coordinates
  $X_A(\s,\tau)$ , ($0 \leq A \leq D-1$) and the world-sheet metric
 $ g_{\a\b}(\s,\tau)$ .

Extremizing  $S$  with respect to them yields the classical
equations of motion:
 \begin{eqnarray}\label{movi}
 \partial^{\a}[\sqrt g\, G_{AB}(X) \; \partial_{\a}X^{B}(\s,\tau)]  &=&
 \frac{1}{2}\, \sqrt g\; \partial_{A}G_{CD}(X) \,  \partial_{\a}X^{C}(\s,\tau)
  \, \partial^{\a}X^{D}(\s,\tau) \\
   & & 0 \le A \le D-1 \cr \label{vincu} \cr
 T_{\a\b} ~ &\equiv&  ~ G_{AB}(X)[ \, \partial_{\a}X^{A}(\s,\tau) \,
  \partial_{\b}X^{B}(\s,\tau) \cr
   & -&   \frac{1}{2} \, g_{\a\b}(\s,\tau) \,  \partial_{\g}X^{A}(\s,\tau) \,
 \partial^{\g}X^{B}(\s,\tau)\, ]= 0 ~~,
  \quad 0 \le \a , \beta \le 1.
\end{eqnarray}
 Eqs. (\ref{vincu})  contain only first derivatives and are therefore a set
  of constraints. Classically, we can always use the reparametrization
 freedom (\ref{reparam}) to recast the world-sheet metric on diagonal form
 \begin{equation}\label{liou}
  g_{\a\b}(\s,\tau)  =  \exp[ \phi(\s,\tau) ]\, \, {\rm diag}( -1, +1)
\end{equation}
 In this conformal gauge, eqs. (\ref{movi}) -  (\ref{vincu}) take the
simpler form:
 \begin{eqnarray}\label{conouno}
 \partial_{-+}X^{A}(\s,\tau)  +   \Gamma^{A}_{BC}(X)\,  \partial_{+}
  X^{B}(\s,\tau) \, \partial_{-}X^{C}(\s,\tau) =  0~, \quad
   0 \le A \le D-1 ,
\end{eqnarray}
\begin{eqnarray} \label{conodos}
         T_{\pm\pm} \equiv G_{AB}(X) \, \partial_{\pm}X^{A}(\s,\tau) \,
\partial_{\pm}X^{B}(\s,\tau)\equiv 0\, , \quad  T_{+-} \equiv T_{-+} \equiv 0
  \end{eqnarray}
 where we introduce light-cone variables  $x_{\pm} \equiv \s \pm \tau $
   on the world-sheet and where  $\Gamma^{A}_{BC}(X)$  stand for the
 connections (Christoffel symbols) associated to the metric  $ G_{AB}(X)$.

Notice that these equations in the conformal gauge are still invariant
under the conformal reparametrizations:
 \begin{equation}\label{conftr}
 \s +  \tau \to  \s'  +  \tau' =  f(\s+ \tau) \qquad , \qquad
\s - \tau \to \s' - \tau'   =g(\s-\tau)
 \end{equation}
Here $f(x)$ and  $g(x)$ are arbitrary functions.

The string boundary conditions in curved spacetimes are identical to
 those in Minkowski spacetime. That is,
 \begin{eqnarray}\label{condc}
  X^{A}(\s + 2 \pi,\tau) = \,   X^{A}(\s,\tau) \quad & {\rm closed
\,strings} \cr \cr
    \partial_{\s}X^{A}(0,\tau)=\, \partial_{\s} X^{A}(\pi,\tau) =  \,0
 \quad &  {\rm open \,strings. }
 \end{eqnarray}

\subsection{A brief review on strings in Minkowski spacetime}

         In flat spacetime eqs.(\ref{conouno})  become linear
\begin{equation}\label{dalam}
 \partial_{-+}X^{A}(\s,\tau)=0~, \quad 0 \le A \le D-1 ,
\end{equation}
and one can solve them
 explicitly as well as the quadratic constraint (\ref{conodos}) [see
below]:
\begin{equation}\label{vinm}
 \left[ \partial_{\pm}X^{0}(\s,\tau) \right]^2 - \sum_{j=1}^{D-1}
\left[ \partial_{\pm}X^{j}(\s,\tau) \right]^2 = 0
\end{equation}

 The solution of eqs.(\ref{dalam})   is
usually written for  closed strings as
\begin{equation}  \label{solM}
         X^{A}(\s,\tau)  =  q^A + 2 p^A \a' \tau + i \sqrt{\a'} \sum_{n \neq 0}
   \frac{1}{n}\{ \a^{A}_{n} \exp[in(\s - \tau)]
         + \tilde  \a^{A}_{n} \exp[-in(\s + \tau )] \}
 \label{solm}
\end{equation}
 where  $q^A$  and  $p^A$  stand for the string center of mass position
  and momentum and  $ \a^{A}_{n}$  and $\tilde \a^{A}_{n}$  describe the right
 and left oscillator modes of the string, respectively. Since the
string coordinates are real,
$$
 {\bar \a}^{A}_{n} =  \a^{A}_{-n} \quad, \quad  {\bar  {\tilde
\a}}^{A}_{n} =
{\tilde \a}^{A}_{-n}
$$
This resolution is no more
 possible in general for curved spacetime where the equations of motion
 (\ref{conouno})  are non-linear. In that case, right and left movers
 interact with themselves and with each other.

In Minkowski spacetime we can also write the solution of the string equations
of motion (\ref{dalam}) in the  form
\begin{equation}
         X^{A}(\s,\tau)  = l^A (\s + \tau) + r^A (\s - \tau)
\end{equation}
where $l^A (x)$ and  $r^A (x)$ are arbitrary functions. Now, making an
appropriate  conformal transformation (\ref{conftr}) we can turn any
of the string coordinates $ X^{A}(\s,\tau)$ (but only one of them)
into a constant times $\tau$. The most convenient choice is the
light-cone gauge where
\begin{equation}\label{gauge}
U \equiv X^0 - X^1 =  2 \, p^U \a' \tau .
\end{equation}
That is, there are no string oscillations along the $U$ direction in
the light-cone gauge. We have still to impose the constraints
(\ref{vinm}). In this gauge they take the form
\begin{equation} \label{V}
 \pm 2 \a' p^U\,
\partial_{\pm}V(\s,\tau)= \sum_{j=2}^{D-1}
\left[ \partial_{\pm}X^{j}(\s,\tau) \right]^2
\end{equation}
where $V \equiv X^0 + X^1$. This shows that $V$ is not an independent
dynamical variable since it expresses in terms of the transverse
coordinates $X^2, \ldots , X^{D-1}$. Only $q^V$ is an independent
quantity.

The physical picture of a string propagating in Minkowski spacetime
clearly emerges in the light-cone gauge. The gauge condition
(\ref{gauge}) tells us that the string `time' $\tau$ is just
proportional to the physical null time $U$. Eqs.(\ref{solM}) shows
that the string moves as a whole with constant speed while it
oscillates around its center of mass. The oscillation frequencies are
all integers multiples of the basic one. The string thus possess an
infinite number of normal modes; $   \a^{A}_{n}, \tilde  \a^{A}_{n}$
classically describe their oscillation amplitudes. Only the modes in
the direction of the transverse coordinates  $X^2, \ldots , X^{D-1}$
are physical. This is intuitively right, since a longitudinal or a
temporal oscillation of a string is meaningless. In summary, the
string in Minkowski spacetime behaves as an extended and composite
relativistic object formed by a
$2(D-2)$-infinite set  of harmonic oscillators.

Integrating eq.(\ref{V}) on $\s$ from $0$ to $2\pi$ and inserting
eq. (\ref{solM}) yields the classical string mass formula:
\begin{equation}\label{masac}
m^2 \equiv  p^U p^V - \sum_{j=2}^{D-1} (p^j)^2 = {1 \o {\a'}}
\sum_{j=2}^{D-1}  \sum_{n=1}^{_\infty}
 \left[   \a^{j}_{n} \, \a^{j}_{-n} + {\tilde  \a^{j}_{n}} \,
{\tilde  \a^{j}_{-n}} \right]
\end{equation}
We explicitly see how the  mass of a string depends on its excitation
state. The classical string spectrum is continuous as we read from
eq.(\ref{masac}). It starts at $m^2 = 0$ for an unexcited string:
$\a^{j}_{n} = \tilde  \a^{j}_{n} = 0 $ for all  $ n$ and $j$.

The independent string variables are:

the transverse amplitudes $\{  \a^{j}_{n},  {\tilde  \a^{j}_{n}},
n\varepsilon {\cal Z}, n \neq 0, \; 2\leq j \leq D-1 \}$,

the transverse center of mass variables $\{ q^j , p^j , \; 2\leq j
\leq D-1 \}$,

$q^V$ and $p^U$.

Up to now we have considered a classical string.

At the quantum level one imposes the canonical commutation relations (CCR)
\begin{eqnarray}\label{ccr}
 [ \a^{i}_{n}, \a^{j}_{m} ] &=& n \; \delta_{n,-m}\, \delta^{i,j} \; ,\cr\cr
 [ {\tilde \a^{i}_{n}, \a^{j}_{m} }] &=& n \; \delta_{n,-m}\, \delta^{i,j}
	\; ,\cr\cr
 [ {\tilde \a^{i}_{n}}, \a^{j}_{m} ] &=& 0 \; ,\cr\cr
 [ q^i , p^j ] = i  \; \delta^{i,j} \quad &,&\quad [q^V, p^U ] = i
\end{eqnarray}
All other commutators being zero. An order prescription is needed to
unambiguously express the different physical operators in terms of
those obeying the CCR. The symmetric ordering is the simplest and more
convenient.

The space of string physical states is the the tensor product of the
Hilbert space of the $D-1$ center of mass variables $q_V, p_U, \{ q^j
, p^j , 2\leq j \leq D-1 \}$, times the Fock space of the harmonic
transverse modes. The string wave function is then the product of a
center of mass part times a harmonic oscillator part. The center of
mass can be taken, for example, as a plane wave. The harmonic oscillator part
can be written as the creation operators $ {\a^{j}_{n}}^{\dag},{\tilde{
{\a}^{j}_{n}}}^{\dag},\; n \geq 1,\;   2\leq j \leq D-1$ acting on the
oscillator ground state $|0>$.  This state is defined as usual by
$$
\a^{j}_{n}\; |0> = {\tilde \a^{j}_{n}}\;|0> = 0, \quad {\rm for~all~}
 n \geq 1,\;   2\leq j \leq D-1
$$
Notice that a string describes {\bf one particle}. The kind of
particle described depends on the oscillator wave function.
The mass and spin can take an infinite number of  different values.
That is, there is  an infinite number of different possibilities   for
the particle described by a string.

Let us consider the quantum mass spectrum. Upon symmetric ordering the
mass operator becomes,
\begin{equation}
m^2= {1 \o {2 \a'}}
\sum_{j=2}^{D-1}  \sum_{n=1}^{_\infty}
 \left[   \a^{j}_{n} \, \a^{j}_{-n} + \a^{j}_{-n} \, \a^{j}_{n}+
{\tilde  \a^{j}_{n}} \, {\tilde  \a^{j}_{-n}}+{\tilde  \a^{j}_{-n}} \,
{\tilde  \a^{j}_{n}} \right] \; .
\end{equation}
Using the commutation rules (\ref{ccr}) yields
\begin{equation}
m^2= {{D-2}\over  {\a'}}      \sum_{n=1}^{_\infty} n +
{1 \o {2 \a'}}\sum_{j=2}^{D-1}  \sum_{n=1}^{_\infty}
 \left[  {\a^{j}_{n}}^{\dag} \, \a^{j}_{n}+
{\tilde  {\a^{j}_{n}}}^{\dag} \,{\tilde  \a^{j}_{n}} \right]
\end{equation}
The divergent sum in the first term can be defined through analytic
continuation of the zeta function
\begin{equation}
\zeta(z) \equiv  \sum_{n=1}^{_\infty} { 1 \over {n^z}}
\end{equation}
One finds  $\zeta(-1) = -1/12$ \cite{grad}. Thus,
\begin{equation}\label{masaq}
m^2= -{{D-2}\over  {12 \a'}}
+ {1 \o {2 \a'}} \sum_{j=2}^{D-1}  \sum_{n=1}^{_\infty}
 \left[  {\a^{j}_{n}}^{\dag} \, \a^{j}_{n}+
{\tilde  {\a^{j}_{n}}}^{\dag} \,{\tilde  \a^{j}_{n}} \right]
\end{equation}
Hence, the string ground
state  $|0>$ has a negative mass squared
\begin{equation}
m_0^2 =  -{{D-2}\over  {12 \a'}}
\end{equation}
Such particles are called tachyons and exhibit unphysical
behaviours. When fermionic degrees of freedom are associated to the
string the ground state becomes massless (superstrings) \cite{gsw}.

Notice that the appearance of a negative mass square yields a
dispersion relation $E^2=p^2 - |m_0^2|$ similar to waves with Jeans
unstabilities \cite{ll}.

Let us consider now excited states.

The constraints (\ref{vinm}) integrated on $\s$ from $0$ to $2\pi$
impose
\begin{equation}
\sum_{j=2}^{D-1}  \sum_{n=1}^{_\infty}
   (\a^{j}_{n})^{\dag} \, \a^{j}_{n} =
\sum_{j=2}^{D-1}  \sum_{n=1}^{_\infty}
{\tilde  {(\a^{j}_{n})}}^{\dag} \,{\tilde  \a^{j}_{n}}
\end{equation}
on the physical states. This means that the number of left and right
modes coincide in all physical states.

The first excited state is then described by
\begin{equation}\label{grav}
|i, j > = {\tilde  {(\a^{i}_{1})}}^{\dag} (\a^{j}_{1})^{\dag} \; |0>
\end{equation}
times the center of mass wave function. We see that this wavefunction
is a symmetric tensor in the space indices $i,j$. It describes
therefore a spin two particle plus a spin zero particle (the trace part).

{}From eqs.(\ref{masaq}-\ref{grav}) follows that
\begin{equation}
m^2\;|i, j > =  -{{D-26}\over  {12 \a'}} |i, j >
\end{equation}
This state is then a massless particle only for $D=26$. In such
critical dimension we have then a graviton (massless spin 2 particle)
and a dilaton  (massless spin 0 particle) as string modes of
excitation. For superstrings the critical dimension turns to be $D=10$.

We shall consider, as usual, that only four space-time dimensions are
uncompactified. That is, we shall consider the strings as living
on the tensor  product of a curved four dimensional space-time with
lorentzian signature and a compact space which is there  to cancel
the anomalies. Therefore, from now on strings will propagate in the
curved (physical) four dimensional space-time. However, we will find
instructive to study the case where this curved space-time has
dimensionality  $D$, where $D$ may be 2, 3 or arbitrary.

\subsection{The string energy-momentum tensor and the string invariant size}

The spacetime string energy-momentum tensor follows (as usual) by taking the
functional derivative of the action (\ref{accion}) with respect to the
metric $ G_{AB} $ at the spacetime point $X$. This yields,

\begin{eqnarray}
\sqrt{-G}~ T^{AB}(X) = \frac{1}{2\pi \alpha'} \int d\sigma d\tau
\left( {\dot X}^A {\dot X}^B -X'^A X'^B \right)
\delta^{(D)}(X - X(\sigma, \tau) )
\label{tens}
\end{eqnarray}
where dot and prime stands for $\partial/\partial\tau$ and
 $\partial/\partial\sigma $, respectively.

Notice that $X$ in eq.(\ref{tens}) is just a spacetime point whereas $
X(\sigma, \tau)$ stands for the string dynamical variables. One sees
from  the Dirac delta in eq.(\ref{tens}) that $ T^{AB}(X) $ vanishes
unless $X$ is exactly on the string world-sheet.
We shall not be interested in the detailed structure of the classical strings.
It is the more useful to integrate
the energy-momentum tensor (\ref{tens}) on a volume
that completely encloses the string. It takes then the  form \cite{ijm}
\begin{equation}
\Theta^{AB}(X^0)  =  \frac{1}{2\pi \alpha'} \int d\s d\tau
\left( {\dot X}^A {\dot X}^B -X'^A X'^B \right)
\delta(X^0 - X^0(\tau,\s ) ).
\end{equation}
When $ X^0 $ depends only on $\tau$, we can easily integrate over
$\tau$ with the result,
\begin{equation}\label{inten}
\Theta^{AB}(X^0)  =  \frac{1}{2\pi \alpha' |{\dot X}^0(\tau)|}
\int_0^{2\pi}d\s \left[ {\dot X}^A {\dot X}^B -X'^A X'^B \right]_
{\tau = \tau(X^0)}
\end{equation}

Another relevant physical magnitude for strings is the invariant
size. We define the invariant string size $ds^2$ using the  metric induced  on
the string world-sheet:
\begin{equation}\label{intinv}
ds^2 = G_{AB}(X) \, dX^A \, dX^B
\end{equation}
Inserting  $ dX^A = \partial_+X^A \; dx^+ +  \partial_-X^A \; dx^- $, into
eq.(\ref{intinv}) and taking into account the constraints
(\ref{conodos}) yields
\begin{equation}\label{tam}
 ds^2 = 2\, G_{AB}(X) \, \partial_+X^A \, \partial_-X^B
\left(d\tau^2 - d\s^2\right) =  G_{AB}(X) \, {\dot X}^A \, {\dot
X}^B \left(d\tau^2 - d\s^2\right) .
\end{equation}
Thus, we define the string size as the integral of $ \sqrt{G_{AB}(X)
 \, {\dot X}^A \, {\dot X}^B}$ over $\s$ at fixed $\tau$.

Let us now consider a circular string as a simple example of a string
solution in Minkowski spacetime.
\begin{eqnarray}\label{anilM}
X^0(\s,\tau) &=& \a' E \tau  \quad ,  \quad X^3(\s,\tau) =  \a' p \tau
\cr & & \cr
X^1(\s,\tau) &= &\a' m \cos\tau \cos\s = {{  \a' m} \over 2} \left[ \cos(\tau
+\s) +  \cos(\tau-\s) \right] \\ & & \cr
X^2(\s,\tau) &=& \a' m \cos\tau \sin\s = {{  \a' m} \over 2} \left[ \sin(\tau
+\s) -  \sin(\tau-\s) \right] \nonumber
\end{eqnarray}
This is obviously a solution of eqs.(\ref{dalam}) where
only the modes $n = \pm 1, j=1,2$ are excited.  The constraints
(\ref{vinm}) yields
$$
E^2 = p^2 + m^2
$$
Eqs.(\ref{anilM}) describe a circular string in the $X^1,X^2$ plane,
centered in the origin and
with an oscillating radius $\rho(\tau) =  \a' m \cos\tau $. In
addition  the string moves uniformly in the $z$-direction with speed
$p/E$. (That is, $p$ is its momentum in the $z$-direction).
The oscillation amplitude $m$ can be identified  with the string mass and $E$
with the string energy. Notice that the string time $\tau$ is here
proportional to the physical time $X^0$ (this solution is not in the
light-cone gauge).

It is instructive to compute the  integrated energy-momentum tensor
(\ref{inten}) for this string solution. We find in the rest frame
($p=0$) that it takes the fluid form

\begin{eqnarray}
\Th_A^B = \left(\begin{array}{rrrr}
\rho & 0 & 0 & 0 \\
0  & -p & 0  & 0 \\
0 & 0  & -p & 0 \\
0  & 0  &  0 & 0 \end{array}\right)
\label{tani}
\end{eqnarray}
 where
\begin{equation}\label{eyp}
\rho = E = m \quad , \quad p = - {m \over 2} \cos(2\tau)
\end{equation}
We see that the total energy coincides with $m$ as one could expect
and that the (space averaged) pressure oscillates around zero. That is the
string pressure goes through positive and negative values. The time
average of $p$ vanishes.  The string behaves then as  cold matter
(massive particles).

The upper value of $p$ equals $E/2$. This is precisely the relation
between $E$ and $p$ for radiation (massless particles). (Notice that
this circular strings lives on a two-dimensional plane). The lower
value of $p$ correspond to the limiting value allowed by the strong
energy condition in General Relativity \cite{hawel}.
We shall see below that these two extreme values of $p$ appear in the
general context of cosmological spacetimes.

The invariant size of the string solution (\ref{anilM}) follows by
inserting eq.(\ref{anilM}) into eq.(\ref{tam}). We find
\begin{equation}
ds^2 = (\a' m)^2  \left(d\tau^2 - d\s^2\right)
\end{equation}
Therefore, this string solution has a constant size $2\pi \a' m$.

One obtains the invariant size for a generic solution in Minkowski
spacetime inserting  the general solution (\ref{solM}) into
$\partial_+X^A \partial_-X_A$. This gives constant plus oscillatory
terms. In any case the  invariant string size is always {\bf bounded}
in Minkowski spacetimes. We shall see how differently behave strings in curved
spacetimes.

\section{\bf String propagation in cosmological spacetimes}

We obtain  in this section  physical string properties
from the exact string solutions in cosmological spacetimes.

We consider strings in spatially homogeneous and isotropic
universes with metric

\begin{equation}
ds^2 = (dT)^2 - R(T)^2 \sum_{i=1}^{D-1}(dX^i)^2
\label{met}
\end{equation}

where $T$ is the cosmic time and the function $R(T)$
is called the scale factor. In terms of the conformal time

\begin{equation}
\eta = \int^{T} \frac{d T}{R(T)}~,
\label{tco}
\end{equation}

the metric (\ref{met}) takes the form

\begin{equation}
ds^2 = R(\eta)^2 \left[ (d\eta)^2 - \sum_{i=1}^{D-1}(dX^i)^2 \right]
\label{metc}
\end{equation}

The classical string equations of motion can be written  here as
\begin{eqnarray}
\partial^2T& - R(T)\; {{{\rm d}R}\over{\,{\rm d}T}}\;
\sum_{i=1}^{D-1}(\partial_\mu X^i)^2=0\,,\label{eqmov}\\
\partial_\mu&\left[R^2 \partial^\mu X^i\right]=0\,,
\qquad 1\leq i\leq D-1\,,\nonumber
\end{eqnarray}
 and the constraints are

\begin{equation}
T_{\pm\pm}=(\partial_\pm T)^2 - R(T)^2 \; (\partial_\pm X^i)^2=0\,.
\label{vinc}
\end{equation}
The most relevants universes corresponds to power type scale factors.
That is,
\begin{equation}
 R(T) = a \; T^{\a} = A \; \eta^{k/2}
\end{equation}
where $ \a = {k \o {k+2}}$. For different values of the exponents we
have either FRW or inflationary universes.
$$
{\rm FRW:~} 0<k \le \infty, \; 0<\a\le 1  = \cases{
	\a = 1/2 , \; k=2, & {\rm radiation dominated} \cr
	\a = 2/3, \; k=4, & {\rm matter dominated} 	\cr
	 \a = 1 , \; k = \infty, &  {\rm `stringy'} \cr}
$$

\begin{equation}
{\rm Inflationary:~} -\infty <k < 0 , \; \a< 0  {\rm ~and~}\a>1
= \cases{
\a =  \infty , \; k=-2, & $R(T) = e^{HT}$,~ {\rm de ~Sitter}, \cr
	\a > 1, \; k<-2 ,  & {\rm power inflation} \cr
 \a < 0 , \; -2 < k < 0 , & {\rm superinflationary} \cr}
\end{equation}

As we will see below, $T^{AB}(X)$ takes the fluid form for string
solutions in cosmological spacetimes,
allowing us to define the string pressure $p$ and energy
density $\rho$ :

\begin{eqnarray}
T_A^B = \left(\begin{array}{rrrr}
\rho & 0 & \cdots & 0 \\
0  & -p & \cdots & 0 \\
\cdots & \cdots  & \cdots & 0 \\
0  & 0  &  \cdots & -p \end{array}\right)
\label{tflu}
\end{eqnarray}

Notice that the continuity equation

\begin{eqnarray}
D^A\;T_A^B = 0 \nonumber
\end{eqnarray}

takes here the form

\begin{equation}
\dot{\rho} + (D-1)\, H\, (p + \rho ) = 0
\label{cont}
\end{equation}

where $ H \equiv {{1}\over {R}} {{dR}\over{dT}} $.

For an equation of state of the type of a perfect fluid, that is

\begin{equation}
p = (\gamma - 1 )\;\rho \qquad , \qquad \gamma = {\rm~constant},
\label{eflu}
\end{equation}

 eqs.(\ref{cont}) and (\ref{eflu}) can be easily integrated with the
result

\begin{equation}
\rho = \rho_0~R^{\,\gamma(1-D)}~~.
\label{ror}
\end{equation}

For $\gamma = 1$ this corresponds to cold matter $(p = 0)$ and
for $\gamma = \frac{D}{D-1} $ this describes radiation with
$p = {{ \rho}\over{D-1}}$.

\subsection{1+1 Dimensional Universes. Straight Strings in
D-Dimensional Universes}

Let us start by considering strings in this simpler case.
For a $D=1+1$ dimensional universe or for a straight string
parallel to the $X$-axis in a $D$-dimensional universe,
 the metric (\ref{met}) takes the form

\begin{eqnarray}
ds^2 = (dT)^2 - R(T)^2 \;(dX)^2
\nonumber
\end{eqnarray}

It is convenient to start by solving the constraints (\ref{vinc})

\begin{equation}
(\partial_\pm T)^2 = R(T)^2 \; (\partial_\pm X)^2 \, .
\label{vincud}
\end{equation}

They reduce to
\begin{equation}
\partial_\pm T = \epsilon_{\pm}~~ R(T) \; \partial_\pm X \, .
\label{raiz}
\end{equation}

where $ \epsilon_{\pm}^2 = 1$ . Using the conformal time (\ref{tco})
, eq.(\ref{raiz}) yields

\begin{eqnarray}
\partial_\pm ( \eta -  \epsilon_{\pm} X ) = 0 \, .
\nonumber
\end{eqnarray}

We find a first family of solutions choosing
 $ \epsilon_{\pm} = \pm 1 $ . Then
\begin{equation}
\eta + X = \phi(\sigma + \tau) ~~,~~ \eta - X = \chi(\sigma - \tau)
\label{sold}
\end{equation}
Where $\phi$ and $\chi$ are arbitrary functions of one variable.
It is easy now to check that eq.(\ref{sold}) fulfills the string
equations of motion (\ref{eqmov}).

The solution (\ref{sold}) is analyzed in detail for de Sitter
spacetime $( R(T) = e^{H T} )$ in ref.\cite{prd} where the
global topology of the space is taken into account.

Since one can always perform conformal transformations
\begin{eqnarray}
\sigma + \tau \to f(\sigma + \tau) \quad , \quad
\sigma - \tau \to g(\sigma - \tau)\quad , \nonumber
\end{eqnarray}

with arbitrary functions $f$ and $g$ , the solution (\ref{sold})
has no degrees of freedom other than topological ones.

Let us compute the energy momentum tensor for the string
solution  (\ref{sold}). We find from eqs.(\ref{tens}) and  (\ref{sold}),

\begin{eqnarray}
\sqrt{-G}~ T^{00}(X) & = & \frac{1}{2\pi \alpha'} \int d\sigma d\tau
\left( {\dot \eta}^2 -{\eta'}^2 \right)
\delta(\eta - \eta(\sigma, \tau) )\delta(X - X(\sigma, \tau) ) \nonumber\\
& = & \frac{1}{2\pi \alpha'} {{ {\dot \eta}^2 -{\eta'}^2}\over J}
\end{eqnarray}

where $ J = \frac{\partial(X,\eta)}{\partial(\sigma,\tau)} $ is the
jacobian. From eq.(\ref{sold}) we find $J = -\chi'\,\phi'$ and $
 {\dot \eta}^2 -{\eta'}^2  =  -\chi'\,\phi'$. Then,
\begin{eqnarray}
\sqrt{-G}~ T^{00}(X) = \frac{1}{2\pi \alpha'} ~.
\nonumber
\end{eqnarray}
 We analogously find $ {\dot X}^2 -{X'}^2  =  -\chi'\,\phi'$. Then
\begin{eqnarray}
\sqrt{-G}~ T^{11}(X) = -\frac{1}{2\pi \alpha'} ~~,~~ T^{01}(X)=0~.
\nonumber
\end{eqnarray}

That implies,
\begin{equation}
\rho = \frac{1}{2\pi \alpha'} ~~,~~ p= -\frac{1}{2\pi \alpha'}~~,~~
p + \rho = 0 .
\label{rop}
\end{equation}

We find a constant energy density and a constant {\bf negative}
pressure.
They exactly fulfill the continuity equation (\ref{cont}).
These results hold for {\bf arbitrary} cosmological spacetimes
in $1 + 1$ dimensions. That is, for  arbitrary factors $R(T)$.
In particular they are valid for strings winded $n$-times around the de Sitter
universe \cite{prd}.

A second family of string solutions follows from eq.(\ref{vincud})
by choosing
\begin{equation}
\eta = \pm X + C_{\pm}
\label{etx}
\end{equation}
where $C_{\pm}$ is a constant.
Then, the string equations of motion (\ref{eqmov})
become
\begin{equation}
\partial_{\mu}\left[R(\eta)^2\;\partial^{\mu} \eta \right]=0
\label{buf}
\end{equation}
Using eq.(\ref{etx}), we find that the energy-momentum tensor
(\ref{tens}) is traceless for this string solution:
\begin{equation}
T^{00} = T^{11} \quad,\quad {\rm that~is~~} p = \rho
\label{duud}
\end{equation}
Let us call
\begin{eqnarray}
V(\eta) = \int^{\eta}R^2(x) dx \nonumber
\end{eqnarray}
Then, eq.(\ref{buf}) implies that
\begin{eqnarray}
\left( {{\partial^2}\over{ \partial \tau ^2}}-
{{\partial^2}\over{ \partial \sigma ^2}} \right) V(\eta) = 0 . \nonumber
\end{eqnarray}
The general solution $\eta = \eta(\sigma, \tau) $ is implicitly
defined by:
\begin{eqnarray}
\int^{\eta}R^2(x) dx = A(\sigma -  \tau) + B(\sigma + \tau)~.\nonumber
\end{eqnarray}
where $A(x)$ and $B(x)$ are arbitrary functions. Upon a conformal
transformation, without loss of generality we can set
\begin{eqnarray}
A(\sigma -  \tau) + B(\sigma + \tau) ~~\Rightarrow ~~\tau~. \nonumber
\end{eqnarray}
Hence, $\eta$ is in general a function solely of
$\tau$ with
\begin{eqnarray}
\tau = \int^{\eta}R^2(x) dx ~~ {\rm or} ~~{{d\eta} \over {d\tau}} =
{1 \over {R^2(\eta)}}~.\nonumber
\end{eqnarray}
The energy-momentum tensor (\ref{tens}) takes for this solution the
form:
\begin{eqnarray}
T_A^B = {1 \over {\alpha'R^2}} \;
\delta(\eta \mp X - C_{\pm})\left(\begin{array}{rr}
1 & 0 \\
0  & -1 \end{array}\right)
\nonumber
\end{eqnarray}
The $\delta(\eta \mp X - C_{\pm})$ characterizes a localized object
propagating on the characteristics at the speed of light.
This solution describes a massless point particle since it has been
possible to gauge out the $\sigma$ dependence .

In summary, the two-dimensional string solutions in cosmological
spacetimes generically obey perfect fluid equations of state with
either
\begin{equation}
p = -\rho ~~(\gamma = 0) \qquad {\rm or}
\qquad p = + \rho ~~(\gamma =2)
\label{prou}
\end{equation}
The respectives energy densities being
\begin{equation}
\rho = \rho_u ~~(\rho_u ={\rm constant)~for~}\gamma = 0\quad {\rm or} \quad
\rho = {{u}\over {R^2}} ~~(u={\rm constant)~for~}\gamma = 2 .
\label{rodu}
\end{equation}
These behaviors fulfill the continuity equation (\ref{cont}) for  $D= 2$ .

\subsection{2+1 Dimensional Universes.  Ring Strings in
D-Dimensional Universes}

A large class of exact solutions describing one string and multistrings
has been found in  $2+1$ dimensional de Sitter universe \cite{dms}
 -\cite{igor}.
For power-like expansion factors $ R(\eta)^2 = A \eta^k ~,~
(k \neq -2) $ only ring solutions are known\cite{din}.
($k = -2 $ corresponds to de Sitter spacetime).
These string solutions are on a two-dimensional plane that can be considered
embedded on a D-Dimensional universe.

Ring solutions correspond to the Ansatz \cite{din}:
\begin{eqnarray}\label{ansatz}
T & = & T(\tau) \nonumber \\
X^1 & = & f(\tau) \cos \sigma \\ \nonumber
X^2 & = & f(\tau) \sin \sigma \label{ani}  \nonumber
\end{eqnarray}
The total energy of one string is then given by (recall $G^{00}=1$)
 \begin{eqnarray}\label{enani}
E(T) = \int d^{D-1}X \sqrt{-G}~T^{00}(X) = {1 \over {\alpha'}}
{{dT} \over {d \tau}}\nonumber
\end{eqnarray}
More generally, the energy-momentum tensor integrated on a volume
that completely encloses the string, takes the form \cite{ijm}
\begin{eqnarray}
\Theta^{AB}(X) & = & \frac{1}{2\pi \alpha'} \int d\sigma d\tau
\left( {\dot X}^A {\dot X}^B -X'^A X'^B \right)
\delta(T - T(\tau) ) \nonumber \\
& = & \frac{1}{2\pi \alpha' |{\dot X}^0(\tau)|}
\int_0^{2\pi}d\sigma \left[ {\dot X}^A {\dot X}^B -X'^A X'^B \right]_
{\tau = \tau(T)}  \nonumber
\end{eqnarray}
For multistring solutions, one must sum over the different roots
$\tau_i$ of the equation $ T = T(\tau) $, for a given $T$.

We find for the ring ansatz eq.(\ref{ani}):
\begin{eqnarray}
\Theta^{00}(X) & = & E(T) \nonumber \\
\Theta^{11}(X) & = & \Theta^{22}(X) =
{1 \over {2 \alpha' |{\dot X}^0(\tau)|}}
[{\dot f}^2 - f^2 ] \nonumber \\\nonumber
\Theta^{01}(X) & = & \Theta^{02}(X) = \Theta^{12}(X) = 0 ~.
\label{tent}\\ \nonumber
\end{eqnarray}
That is,

\begin{eqnarray}
\Theta_A^B = \left(\begin{array}{rrr}
E & 0  & 0 \\
0  & -P  & 0 \\
0  & 0   & -P \end{array}\right)
\nonumber
\end{eqnarray}

where,
\begin{equation}
E = {1 \over {\alpha'}}~{\dot T}(\tau) \quad,\quad
P = {{R(\tau)^2} \over {2 \alpha' |{\dot X}^0(\tau)|}}~~
[{\dot f}^2 - f^2 ]
\label{epa}
\end{equation}

Let us first consider ring strings in de Sitter universe
\cite{dms} - \cite{dls}. We find there  three different
asymptotic behaviors: stable, unstable and its dual.
The unstable regime appears for $ \eta \simeq {{\tau}\over H}
\to 0 $. That is, when the scale factor tends to infinity.
{}From eqs.(\ref{epa}) and ref.\cite{dms}, we find
\begin{eqnarray}
E(\tau) &\buildrel{\tau \to 0^-}\over=& {1\over {\alpha'H }}
\left( {1 \over {|\tau|}} + 1 \right) \simeq {1\over {\alpha'H }}
\left[ R(\tau) + 1 \right] \to +\infty \nonumber \\
P(\tau) &\buildrel{\tau \to 0^-}\over= & -{1\over {2 \alpha 'H |\tau|}}
\simeq -{{R(\tau)}\over {2 \alpha'H }}\to -\infty \label{desu}
 \nonumber
\end{eqnarray}
(Here $H$ stands for the Hubble constant).
The invariant string size grows as $R/H$ in this unstable regime.
Notice that the pressure is {\bf negative} for the unstable strings
and proportional to the expansion factor $R$.  In this
regime we also see that
\begin{equation}
P   \buildrel{R \to \infty}\over= -{E \over 2} +
{1 \over {2 \alpha' H}} + O(1/R)\to -\infty
\label{peun}
\end{equation}
with $E =  {1\over {\alpha'H }} \left[ R(\tau) + 1 \right] $.
The regime dual to the unstable regime appears when the conformal time $\eta$
tends to infinity. For the solution  $q_-(\sigma,\tau)$ of ref.\cite{dms}
in de Sitter universe, $\eta$ diverges for finite $\tau \to \tau_0$ as
\begin{eqnarray}
\eta   \buildrel{\tau \to \tau_0}\over= {{6\;e^{-\tau_0} \over
{\tau -\tau_0}}} + O(1) \to +\infty \nonumber
\end{eqnarray}
Here
\begin{eqnarray}
\tanh{{\tau_0}\over{\sqrt{2}}} = {1 \over {\sqrt{2}}} \quad,\quad
\tau_0 = \sqrt{2}\,\log(1+\sqrt{2})= 1.246450.... \nonumber
\end{eqnarray}
Then,
\begin{eqnarray}
R(\tau) & \buildrel{\tau \to \tau_0}\over= & {{e^{\tau_0}} \over 6}\,
(\tau -\tau_0) \to 0^+ \nonumber \\
E(\tau) &\buildrel{\tau \to \tau_0}\over=& {1\over {\alpha'H (\tau -\tau_0)}}
= { 0.5796... \over {\alpha'H R}} \to +\infty \nonumber \\
P(\tau) &\buildrel{\tau \to \tau_0}\over=& {1\over {2\alpha'H (\tau -\tau_0)}}
= E/2 \to +\infty \label{dual}   \nonumber
\end{eqnarray}
We call dual to this regime since it appears related to the unstable regime
(\ref{desu}) through the exchange $ R \leftrightarrow 1/R $ .
The invariant string size tends to $ {1\over  H }$ in this regime.

In the stable regime, $\tau \to \infty$,
(and the cosmic time $ T \simeq {{\tau}\over H} \to \infty $),
from eq.(\ref{epa}) and ref.\cite{dms}, we find
\begin{equation}
E(\tau) \buildrel{\tau \to \infty}\over= {1\over {\alpha'H }}
\quad , \quad
P(\tau) \buildrel{\tau \to \infty}\over= \frac{1+\sqrt{2}}{2 H}
e^{-\tau\sqrt{2}} \to 0 .
\label{esta}
\end{equation}

For stable strings in de Sitter universe, the pressure
 is positive and vanishes asymptotically, and the invariant string size
tends to $ {1\over {\sqrt{2} H }}$.

The solution  $q_-(\sigma,\tau)$ in ref.\cite{dms} describes
two ring strings:

a stable string for $ q_0 \to +\infty  \;
( \tau \to \infty ) $
($q_0$ being the hyperboloid time-like coordinate) and

a unstable one for
 $q_0 \to -\infty \; ( \tau \to \tau_0 )$.
 The pressure $P$ depends on $\tau$ ; it is
negative for $ q_0 < l$ and positive for $q_0 > l$, where
$l = -1.385145...$.

It can be noticed that the behaviour eq.(\ref{desu}) for the energy can
be interpreted as an unstable piece $ R/(\alpha' H) $ {\it plus} a
stable one $ 1/(\alpha' H) $ . The constant term   $ 1/(\alpha' H) $
is precisely  the energy for the stable solution eq.(\ref{esta}).

\bigskip

Let us now study the energy and pressure for the ring string solutions
in power-type inflationary universes considered in
\cite{din}. In terms of the conformal time $\eta$ , we have as expansion
factor $ R^2(\eta) = A \, \eta^k $ with $k<0$.

Near $\eta = 0$ , two types of behavior were found \cite{din}.
The first one is a linear behavior
\begin{eqnarray}
\eta  & \buildrel{\tau \to \tau_0}\over= & \tau -\tau_0 +
O(\tau-\tau_0)^2 \nonumber \\\label{etan}
f(\tau) & \buildrel{\tau \to \tau_0}\over= &
1 - {{(\tau-\tau_0)^2}\over{2(k+1)}} ~~{\rm for~} k < -1  \\
{\rm and~} f(\tau) & \buildrel{\tau \to \tau_0}\over= &
1 + c ~(\tau-\tau_0)^{1-k}  ~~{\rm for~} -1<k<0.\nonumber
\end{eqnarray}
Eqs.(\ref{etan}) describe a  expanding string for $ k < 0~
 $ with invariant size $ \simeq (\tau-\tau_0)^{k/2} $.
That is, in inflationary universes
$(k < 0)$ , the string size grows indefinitely for
 $\eta \simeq \tau-\tau_0 \to 0 $ as the universe radius
$R \simeq (\tau-\tau_0)^{k/2} \to +\infty $. The growing of the proper
string size for $\eta  \simeq \tau -\tau_0 \to 0 $ is a typical
feature of string unstability.

Using eqs.(\ref{epa}) and (\ref{etan}), the energy and pressure take the
form,
\begin{eqnarray}
E(\tau) & \buildrel{\tau \to \tau_0}\over= &
{{\sqrt{A}}\over  {\alpha'}}~(\tau-\tau_0)^{k/2}= R/\alpha'
\quad , \nonumber \\
P(\tau) & \buildrel{\tau \to \tau_0}\over= &
-{{\sqrt{A}}\over  {2\alpha'}}~(\tau-\tau_0)^{k/2}= -R/(2 \alpha')
\quad . \label{epun}
\end{eqnarray}
That yields
\begin{equation}
P \buildrel{\tau \to \tau_0}\over= - E/2
\label{eqes}
\end{equation}

Eq.(\ref{eqes}) is also valid  in de Sitter universe for unstable
strings [eq.(\ref{peun})].
In all these cases strings exhibit {\bf negative} pressure with an  equation of
state $P= -E/2$. This equation of state exactly saturates the strong
energy condition in general relativity. This  {\bf unstable} string
behavior  dominates in  all inflationary universes for $R \to \infty$.

The second behavior present near $\eta = 0$ is \cite{din}
\begin{equation}
\eta   \buildrel{\tau \to \tau_0}\over=  (\tau -\tau_0)^{1/(k+1)}
\quad , \quad
f(\tau) \buildrel{\tau \to \tau_0}\over= f_0 \pm  (\tau -\tau_0)^{1/(k+1)}
\label{eand}
\end{equation}
where $f_0$ must be set equal to zero for $-1 < k < 0 $.
The invariant string size behaves for $ \tau \to \tau_0 $ as,
\begin{eqnarray}
S(\tau) & \buildrel{\tau \to \tau_0}\over=  &
\sqrt A~(\tau -\tau_0)^{{k+2}\over {2(k+1)}}  \buildrel{\tau \to
\tau_0}\over= A\; T  \quad {\rm for~} k< 0
\nonumber
\end{eqnarray}
This solution describes a string that collapses for
 $k < -2$ (power inflation) and blows up for $-2<k<0$ (super
inflation). In both cases the string size is proportional to the
horizon size which is of order $T$ (cosmic time).

Here, the expansion factor tends to zero as
\begin{eqnarray}
R(\tau) \buildrel{\tau \to \tau_0}\over=
\sqrt{A}~(\tau -\tau_0)^{{k}\over {2(k+1)}} \to 0.\nonumber
\end{eqnarray}
when $k < -1$ and blows up for $0<k<-1$.

{F}rom eq.(\ref{epa}) we find that
\begin{eqnarray}
P = {A \over {2(k+1)\alpha' R}}\to +\infty \quad , \quad
E = {A \over {(k+1)\alpha' R}} \to +\infty \nonumber
\end{eqnarray}
That is,
\begin{equation}
P \buildrel{\tau \to \tau_0}\over= E/2
\label{duat}
\end{equation}
Notice that the pressure is here {\bf positive}.
This second behaviour is related by duality ($ R \leftrightarrow 1/R $)
to the first  behaviour described by eqs.(\ref{epun})-(\ref{eqes}).
This  is the {\bf dual to unstable} regime. In such regime the
strings behave as massless particles (radiation).

\bigskip

Let us now consider  ring string solutions
in FRW  universes considered in
\cite{din}. These solutions start ex-nihilo at the big bang ($R=0$).
In terms of the conformal time $\eta$ , we have as expansion
factor $ R^2(\eta) = A \, \eta^k $ with $k>0$.

Near $\eta = 0$ , two types of behavior were found \cite{din}.
The first one is a linear behavior
\begin{eqnarray}\label{etanF}
\eta  & \buildrel{\tau \to \tau_0}\over= & \tau -\tau_0 +
O(\tau-\tau_0)^2 \nonumber \\
f(\tau) & \buildrel{\tau \to \tau_0}\over= &
1 - {{(\tau-\tau_0)^2}\over{2(k+1)}} \quad .
\end{eqnarray}
Eqs.(\ref{etanF}) describe a collapsing  string for $ k > 0
 $ with invariant size $ \simeq (\tau-\tau_0)^{k/2} $.
That is, in  FRW universes $(k>0)$, the string size goes to
zero for $\eta \simeq \tau-\tau_0 \to 0 $ as the universe radius
$R \simeq (\tau-\tau_0)^{k/2} \to 0 $.

Using eqs.(\ref{epa}) and (\ref{etanF}), the energy and pressure take the
form,
\begin{eqnarray}
E(\tau) & \buildrel{\tau \to \tau_0}\over= &
{{\sqrt{A}}\over  {\alpha'}}~(\tau-\tau_0)^{k/2}= R/\alpha'
\quad , \nonumber \\
P(\tau) & \buildrel{\tau \to \tau_0}\over= &
-{{\sqrt{A}}\over  {2\alpha'}}~(\tau-\tau_0)^{k/2}= -R/(2 \alpha')
\quad . \label{epunF}
\end{eqnarray}
That yields
\begin{equation}
P \buildrel{\tau \to \tau_0}\over= - E/2
\label{eqesF}
\end{equation}
As we shall see below,  this behaviour is subdominant in   FRW universes.
[Notice that eq.(\ref{eqesF}) is identical  eq.(\ref{eqes}) that holds
 for $R\to\infty$ in power-like inflationary universes ($k<0$)].

The second behavior present near $\eta = 0$ is \cite{din}
\begin{equation}
\eta   \buildrel{\tau \to \tau_0}\over=  (\tau -\tau_0)^{1/(k+1)}
\quad , \quad
f(\tau) \buildrel{\tau \to \tau_0}\over= f_0 \pm  (\tau -\tau_0)^{1/(k+1)}
\label{eandF}
\end{equation}
The invariant string size behaves for $ \tau \to \tau_0 $ as,
\begin{eqnarray}
S(\tau) & \buildrel{\tau \to \tau_0}\over=  &
\sqrt A~(\tau -\tau_0)^{{k}\over {2(k+1)}}\to 0  \quad {\rm for~} k> 0
\nonumber
\end{eqnarray}
This solution describes a string that collapses for $k > 0$.

Here, the expansion factor tends to zero as
\begin{eqnarray}
R(\tau) \buildrel{\tau \to \tau_0}\over=
\sqrt{A}~(\tau -\tau_0)^{{k}\over {2(k+1)}} \to 0.\nonumber
\end{eqnarray}
since $k > 0$.

{F}rom eq.(\ref{epa}) we find that
\begin{eqnarray}\label{epdF}
P = {A \over {2(k+1)\alpha' R}}\to +\infty \quad , \quad
E = {A \over {(k+1)\alpha' R}} \to +\infty \nonumber
\end{eqnarray}
That is,
\begin{equation}
P \buildrel{\tau \to \tau_0}\over= E/2
\label{duatF}
\end{equation}
Notice that the pressure is here {\bf positive}.
This second behaviour is related by duality ($ R \leftrightarrow 1/R $)
to the first  behaviour described by eqs.(\ref{epunF})-(\ref{eqesF}).

Hence, we see  from eqs.(\ref{epunF}-\ref{epdF})
that near the big bang ($R\to 0$)
the {\bf dual} behavior dominates  over the unstable behavior.

It should also be noticed that for this dual behavior, the energy
redshifts exactly as $1/R$ for $R\to 0$.

\bigskip

For large $\tau$, the ring strings exhibit a stable
behaviour\cite{din},
\begin{eqnarray}
\eta(\tau)&\buildrel{{\tau\to+\infty}}\over=
&\tau^{2/(k+2)}\,,\nonumber \\
T(\tau)&\buildrel{{\tau\to+\infty}}\over= &
{{2\sqrt{A}}\over{k+2}}\;\tau ~, \nonumber \\
f(\tau)&\buildrel{{\tau\to+\infty}}\over= &{2\over{k+2}} \; \tau^{-k/(k+2)}
\cos(\tau+\varphi)\,,\label{solasim}\\
\nonumber
\end{eqnarray}
where $ \varphi $ is a constant phase and the oscillation amplitude has been
 normalized. For large $\tau $, the energy and pressure of the
solution (\ref{solasim}) behave as
\begin{eqnarray}
E(\tau) & \buildrel{{\tau\to+\infty}}\over= & {{2\sqrt{A} } \over {\alpha'
(k+2)}} =  {\rm constant~} \quad, \nonumber \\
P (\tau) & \buildrel{{\tau\to+\infty}}\over= & -{{\sqrt{A} } \over {\alpha'
(k+2)}} \cos(2\tau + 2\varphi) \to 0 \quad .
\label{epes}
\end{eqnarray}

This is the analog of the stable behaviour (\ref{esta}) in de Sitter
universe for $ \tau \to \infty $.
Notice that eqs.(\ref{epes}) hold both for  FRW
$( k > 0 )$ and inflationary spacetimes $(k < 0)$.

The behaviour described by eqs.(\ref{epes}) is similar to Minkowski
spacetime. Notice that the factor $  \tau^{-k/(k+2)} $ in the comoving
radius [eq.(\ref{epes})] is just the scale factor. That is, the
physical string radius oscillates with constant amplitude as in
Minkowski spacetime. The fact that the spacetime curvature
tends to zero (except for de Sitter universe) when $T(\tau) \to \infty$
explains here the presence of a Minkowski-type behaviour.

\bigskip

For power-like inflationary universes with $ k < -1 $ a special exact
ring solution exist \cite{din} with
\begin{eqnarray}
\eta = C \, \exp{{{\tau} \over {\sqrt{-k-1}}}} \quad , \quad
f(\tau) = { C \over {\sqrt{-k}}} \exp{{{\tau} \over {\sqrt{-k-1}}}}
\label{exot}
\end{eqnarray}
where $ C $ is an arbitrary constant. For $ k = - 2 $ (de Sitter
universe) this is the solution $q^{(o)}(\sigma, \tau)$ in
ref.\cite{dms} which has constant string size.

For this solution we find
\begin{eqnarray}
E & = & \frac{1}{\alpha'}\sqrt{{A \over {-k-1}}}\;
\exp\left[{{{\tau (k+2)} \over {2\sqrt{-k-1}}}}\right] = K \; R^{1 + 2/k}
\nonumber \\
P & = & -({1 \over 2}+{1 \over k})~E \label{eexo} \nonumber
\end{eqnarray}
where $ K $ is a constant.

This is a fluid-like equation of state with
$\gamma = {1 \over 2}-{1 \over k}$ . Notice that ${1 \over 2} < \gamma
< {3 \over 2} $. For this solution, the energy grows with $R$ as $R$ to
a power $ 1 + 2/k $ where, since $ k < - 1 $ ,
\begin{eqnarray}
- 1 < 1 + {2 \over k} < 1 . \nonumber
\end{eqnarray}
($E$ and $P$ of this solution are constants in de Sitter spacetime $[k
= -2$]).
This means that these special strings are subdominant both for $R \to \infty$
and for $ R \to 0 $ where the unstable strings ( $ E \simeq R $ )
and their dual  ($ E \simeq R^{-1} $)  dominate
respectively.

\bigskip

In ref.\cite{dls}  the exact general evolution of circular strings
in $2+1$ dimensional
de Sitter spacetime is described closely and completely in terms of elliptic
functions. Such solutions follow the form of  eq.(\ref{ansatz}).
The evolution depends on a constant parameter $b$, related to the
string energy, and falls into three classes depending on whether $b<1/4$
(oscillatory motion), $b=1/4$ (degenerated, hyperbolic motion) or
$b>1/4$ (unbounded motion). The novel feature here is that one single
world-sheet generically describes {\it infinitely many} (different and
independent) strings. The world-sheet time $\tau$ is an infinite-valued
function of the string physical time, each branch yields a different
string. This has no analogue in flat spacetime.

We computed in ref.\cite{dls}  the string
energy $E$ as a function of the string proper size $S$, and analyze it
for the expanding and oscillating strings. For expanding strings
$(\dot{S}>0)$: $E\neq 0$ even at $S=0$, $E$ decreases for small $S$ and
increases $\propto\hspace*{-1mm}S$ for large $S$. For an oscillating string
$(0\leq S\leq S_{max})$, the average energy $<E>$ over one oscillation
period is expressed as a function of $S_{max}$ as a complete elliptic
integral of the third kind \cite{dls}.

Let us briefly describe here the periodic solutions found in  ref.\cite{dls}.
The scale factor takes the form
\begin{eqnarray}
R(T)=&
e^{HT_+(\tau)}=e^{\tau [{k\over{1+k^2}}+\Omega\frac{\vartheta'_4}{\vartheta_4}
 ( \Omega y) ]} \;
\frac{\Omega\vartheta'_1(0)}{\pi\vartheta_1(\Omega y)}\;
\mid \frac{\vartheta_4(\Omega(\tau-y))}{\vartheta_4(\Omega\tau)}\mid
\end{eqnarray}
where $0 \leq k \leq 1$ stands for the elliptic modulus,
$$
\Omega \equiv {\pi \o {2 \, \sqrt{k^2 +1} \, K(k)}} \quad , \quad
y \equiv \sqrt{k^2 +1} \; F({ 1 \o \sqrt{k^2 +1}}, k )
$$
and $F(\phi,k)$ stands for a 1st. class elliptic integral.
Notice that $ 1\geq \Omega \geq 0$ for  $0 \leq k \leq 1$.
$b$ relates with $k$ through $\sqrt{b} = {k \o {1 +k^2}}$.
We see that $R(T)$ is an oscillatory function of $\tau$.

The cosmic time can be explicitly written as a Fourier series:
\begin{eqnarray}\label{Tper}
HT(\tau)=\tau\; [{k\over{1+k^2}}+
\Omega\frac{\vartheta'_4}{\vartheta_4}(\Omega y)]+
\log\mid\frac{\Omega\vartheta'_1(0)}{\pi\vartheta_1(\Omega y)}\mid \nonumber
\cr
-4\, \sum_{n=1}^{\infty}\frac{q^n}{n(1-q^{2n})}\sin (n y \Omega)
\sin\left[n\pi\Omega(2\tau-y)\right].
\end{eqnarray}
The comoving string radius $f(\tau)$ [see eq.(\ref{ansatz})] takes here the
form
\begin{eqnarray}\label{fper}
f(\tau) = e^{-H T(\tau)} \; {k\over{\sqrt{1+k^2}}} \; {\rm sn}\left({\tau
\over {\sqrt{1+k^2}}}, k\right) \cr
=\frac{2\pi}{K(k)\sqrt{1+k^2}}
\sum_{n=1}^{\infty}\frac{q^{n-1/2}}{1-q^{2n-1}}
\sin\left[(2n-1)\Omega\tau\right].
\end{eqnarray}

The cosmic time and the comoving string radius are  therefore
completely regular functions of $\tau$, and it follows that
this string solution which is {\it oscillating regularly}
as a function of world-sheet
time $\tau$, is also oscillating regularly when expressed in terms of
hyperboloid time or cosmic time. This solution represents one {\it stable}
string.

The invariant string size results:
\begin{equation}\label{sper}
S(\tau) =  {k\over{H \sqrt{1+k^2}}} \; {\rm sn}\left({\tau
\over {\sqrt{1+k^2}}}, k\right)
\end{equation}
We see that the string oscillates {\bf inside} the horizon:
$$
0 < S(\tau) < \frac{1}{H}
$$

We see from eqs. (\ref{Tper}-(\ref{fper}) that the string oscillations
do not follow a
pure harmonic motion as in flat Minkowski spacetime, but they are
precise superpositions of all frequencies $(2n-1)\Omega$ ($n=1,2,...,\infty$)
with uniquely defined coefficients. The non-linearity of the string equations
in de Sitter spacetime fixes the relation between the mode coefficients.
In the present case the basic frequency $\Omega$ depends on the
string energy, while in Minkowski spacetime the frequencies are
merely $n$.

The energy (\ref{enani}) of this solution is given by\cite{dls}:
\begin{eqnarray}\label{Eper}
E=\frac{1}{H\alpha'}
[{k\over{1+k^2}}+  {1\over{\sqrt{1+k^2}}} \;{\rm zn}({y\o{
\sqrt{1+k^2}}},k)]\;+\;
{\mbox oscillating\;terms},
\end{eqnarray}
where zn$(x,k)$ stands for the Jacobi zeta-function \cite{grad}.
Averaging over a period $2K\sqrt{1+k^2}$, the average energy
$<E>$ is just  the square bracket term.

 The  pressure (\ref{enani}) averaged over a period can be
expressed here as a
combination of complete elliptic integrals of the third
kind. Numerical evaluation gives zero with high accuracy. We have
checked in addition  that the first orders in the expansion in the
 $k$ identically vanish.
We conclude that the average pressure
is zero as in Minkowski spacetime. Thus, in average the
{\it oscillating strings} describe {\it cold matter}.

In the $k \to 0 $ limit the amplitude of the string size
oscillation and the energy goes to zero [see
eq.(\ref{sper}-\ref{Eper})]. Actually the solution disappears in such limit.

The other
interesting limit is $k \rightarrow 1$ where the elliptic solutions turn
into hyperbolic solutions. Then $<E>\rightarrow 1/(H\alpha')$.

More generally a numerical analysis shows that $<E>$ is a
monotonically increasing function of
$b$ for $b\in [0,1/4]$.

In summary, three asymptotic behaviors are exhibited by ring
solutions.

\bigskip

i) Unstable for $R \to \infty$ and $R \to 0$:

$ E = {\rm (constant)~} R \to +\infty ~
,~ P = -E/2 \to -\infty,~S \simeq R \to \infty $.

\medskip

ii) Dual to unstable for $R \to 0$ and  $R \to \infty$:

$E = {\rm (constant)~} R^{-1} \to
+\infty ~,~ P = +E/2 \to +\infty , \\ ~
S \simeq R \to 0$ (except for de Sitter
where $S \to {1 \over H}$).

\medskip

iii) Stable for $R \to \infty:$

$E = {\rm constant~} ,~ P = 0, ~
 S = {\rm constant~} $.

\bigskip

Recall that the unstable behaviour dominates for  $R \to \infty$ in
inflationary universes whereas the dual behaviour dominates in FRW
universes for  $R \to 0$.

\bigskip

In addition we have the special behavior (\ref{eexo}) for $ k < -1 $.
Notice that the three behaviors i)-iii) appear for all expansion factors
$R(\eta)$ . The behaviours i) and ii) are related by the duality
transformation $ R \leftrightarrow 1/R $, the case iii) being invariant
under duality . In the three cases we find perfect fluid equations
[see eq.(\ref{eflu})] with different $ \gamma $ :
\begin{eqnarray}
\gamma_u = 1/2 ~~,~~ \gamma_d = 3/2 ~~,~~ \gamma_s = 1~~, \nonumber
\end{eqnarray}
where the indices $u, d$ and $s$  stand  for `unstable',
`dual' and `stable', respectively.
Assuming a perfect gas of strings on a volume $R^2$ , the energy
density $\rho$ will be proportional to $ E/R^2 $. This yields
the following scaling with the expansion factor using
the energies from i)-iii):
\begin{equation}
\rho_u = {\rm constant~} R^{-1} ~~,~~ \rho_d =  {\rm constant~} R^{-3} ~~,~~
\rho_s =  {\rm constant~} R^{-2} .
\label{rodd}
\end{equation}
All three densities and pressures obey the continuity equation
(\ref{cont}), as it must be.

The factor $1/2$ in the relation between $P$ and $E$ for the  cases i) and
ii) is purely geometric. Notice that this factor was one in $1+1$
dimensions [eq.(\ref{prou})].

\subsection{D-Dimensional Universes}

The solutions investigated in secs. II.A and II.B  can exist for any
dimensionality of the spacetime. Embedded in D-dimensional universes,
the $(1+1)$ solutions of section  II.A describe
straight strings, the $(2+1)$ solutions of section  II.B are circular rings.
 In D-dimensional spacetime, one expects string solutions spread in
$D-1$ spatial dimensions. Their treatment has been done
asymptotically in ref.\cite{gsv}. One finds for $\eta \simeq \tau -
\tau_0 \to 0$ ,  $ R \to \infty$,
\begin{equation}
P_u = -{1 \over {D-1}}\;\rho_u ~~~,~~~  \gamma_u = {{D-2}\over{D-1}}  ~~~
{\rm for~unstable~strings.}
\end{equation}
This relation coincides with eq.(\ref{rop}) for $D = 2$ and with
eqs.(\ref{peun}) and (\ref{eqes}) for $D = 3$.

The energy density scales with $R$ as
\begin{eqnarray}
\rho_u = u \; R^{2-D} ~~~,~~~ R \to \infty
\label{estin}
\end{eqnarray}
(where $u$ is a constant)
in accordance with eq.(\ref{rodu}) for $ D = 2 $ and with
eq.(\ref{rodd}) for $ D = 3 $.

For the dual regime, $ R \to 0$, we have:
\begin{equation}
P_d = +{1 \over {D-1}}\;\rho_d ~~,~~ \gamma_d = {D\over{D-1}} ~~
{\rm with~} \rho_d = d \; R^{-D}~~,~~ R \to 0 ~~,
\label{estdu}
\end{equation}
(where $d$ is a constant).
Eq.(\ref{estdu}) reduces to eq.(\ref{duud}) for $ D = 2 $  and to
eqs.(\ref{dual}) and (\ref{duat}) for $ D = 3 $. In this
dual regime strings have the same equation of state as massless radiation.

Finally, for the stable regime we have
\begin{equation}
P_s = 0 ~~,~~ \gamma_s = 1~~{\rm with~} \rho_s = s \; R^{1-D}
\end{equation}
(where $s$ is a constant).
This regime is absent in $ D = 2 $ and appears for  $ D = 3 $ solutions
in eqs.(\ref{esta}) and (\ref{epes}).
The lack of string transverse modes in $ D = 2 $ explains the absence
of the stable regime there.
The equation of state
for stable strings coincides with the one for cold matter.

In conclusion, an ideal gas of classical strings in cosmological universes
exhibit three different thermodynamical behaviours,
all of perfect fluid type:

\medskip

1) Unstable strings :
negative pressure gas with $\gamma_u = {{D-2}\over{D-1}} $

2) Dual behaviour : positive pressure gas similar to radiation ,
$\gamma_d = {D\over{D-1}} $

3) Stable strings : positive pressure gas similar to cold
matter, $ \gamma_s = 1$.

\medskip

Tables I and II summarize the main string properties for any scale
factor $ R(X^0) $.
\newpage
\begin{centerline}
 {TABLE 1. String energy and pressure as obtained from exact}
 {string solutions for various expansion factors  $R(X^0)$.}

\bigskip

\bigskip

{\bf STRING PROPERTIES FOR ARBITRARY $R(X^0)$}

\end{centerline}
\vskip 20pt
\begin{tabular}{|l|l|l|l|}\hline
$ $&  $ $ & $ $ &  Equation  of State: \\
$ $& $~~$ Energy  & $~~~~$Pressure  & $ $   \\
$ $&  $ $ & $ $ & $~~~p = (\gamma-1) \rho $\\ \hline
D = 1 + 1: two families  & $ $ & $ $ & $ $ \\
\hspace{13mm} of solutions  & $ $ & $ $ & $ $\\ \hline
$ $ & $ $ & $ $ & $ $ \\
(i)$~\eta\pm X=f_{\pm}(\sigma\pm\tau)$  & $~~~E = ~ u \; R $ &
 $~~~~P = - E$ & $~~~~\gamma = 0 $ \\
 (ii)$~\eta\pm X=$~constant & $~~~E  = d /R$ &
$~~~~P = + E$ & $~~~~\gamma = 2 $ \\
$ $ & $ $ & $ $ & $ $ \\ \hline
$D = 2 + 1$: Ring Solutions,  & $ $ & $ $ & $ $ \\
three asymptotic &  $  E = {1 \over {\alpha'}}~{\dot X}^0(\tau)$
& $ P = {{R(\tau)^2} \over {2 \alpha' |{\dot X}^0(\tau)|}}\,
[{\dot f}^2 - f^2 ]  $ & $ $ \\
behaviours (u, d, s) & $ $ & $ $ & $ $ \\ \hline
$ $ & $ $ & $ $ & $ $ \\
(i) unstable for $R\to\infty$ & $E_u  \buildrel{R \to \infty}\over = u
\; R \to \infty$ & $P_u = -E_u/2\to-\infty$ & $~~\gamma_u = 1/2 $ \\
(ii) dual to (i) for $R \to 0$ & $E_d \buildrel{R \to 0}\over = d /R
\to \infty$ & $P_d = + E_d/2 \to \infty$ & $~~\gamma_d = 3/2 $ \\
(iii) stable for  $R\to\infty$ & $E_s = \;$ constant & $P_s = 0 $ &
 $~~\gamma_s = 1 $\\
$ $&  $ $ & $ $ & $ $\\ \hline
D-Dimensional spacetimes: &  $ $ & $ $ & $ $ \\
general asymptotic behaviour & $ $ & $ $ & $ $ \\ \hline
$ $ & $ $ & $ $ & $ $ \\
(i) unstable for $R\to\infty$ & $E_u  \buildrel{R \to \infty}\over = u
\; R \to \infty$ & $P_u = -{E_u\over{D-1}}\to-\infty$ &
$\gamma_u =   \frac{D-2}{D-1} $ \\
(ii) dual to (i) for $R \to 0$ & $E_d \buildrel{R \to 0}\over = d /R
\to \infty$ & $P_d = + {E_d\over{D-1}} \to \infty$ & $\gamma_d = \frac{D}{D-1}
$ \\
(iii) stable for  $R\to\infty$ & $E_s = \;$ constant & $P_s = 0 $ &
 $~~\gamma_s = 1 $\\
$ $&  $ $ & $ $ & $ $\\ \hline
\end{tabular}
\vskip 20pt
\newpage
\newpage
\begin{centerline}
 {TABLE 2. The string energy density and pressure}
 {for a gas of strings can be summarized by the formulas}

{below which become exact for $R \to 0$ and for $R \to \infty$.}

\bigskip

\bigskip

{\bf STRING ENERGY DENSITY AND PRESSURE}

{\bf  FOR ARBITRARY $R(X^0)$}


\end{centerline}
\vskip 20pt
\begin{tabular}{|l|l|l|}\hline
$ $& Energy density: $~\rho \equiv E/R^{D-1}$ & \hspace{10mm}Pressure  \\
$ $&  $ $ & $ $ \\ \hline

Qualitatively correct &  $ $ & $ $\\
formulas for all R and D &
$ ~~\rho = \left( u \; R + {{d} \over R} + s \right) {1 \over
{R^{D-1}}} $ &
$ p  = {1 \over {D-1}} \left( {d \over R} -  u \; R\right) {1 \over
{R^{D-1}}} $ \\
$ $&  $ $ & $ $ \\ \hline
\end{tabular}
\vskip 30pt
\begin{centerline}
 {TABLE 3. The {\bf self-consistent} cosmological solution}
{ of the Einstein equations in General Relativity}

{ with the string gas as source.}

\bigskip
\bigskip

{\bf STRING COSMOLOGY IN GENERAL RELATIVITY}


\end{centerline}
\vskip 20pt
\begin{tabular}{|l|l|l|}\hline
Einstein equations & \hspace{4mm}Expansion factor  &\hspace{6mm} Temperature \\
(no dilaton field) & \hspace{12mm} $R(X^0)$ &  \hspace{13mm}$T(R)$ \\
\hline
$ $&  $ $ & $ $ \\
$ X^0 \to 0 $ & ${D \over 2}
\left[{{2 d}\over {(D-1)(D-2)}}\right]^{1 \over D}~ (X^0)^{2 \over D}$&
\hspace{4mm}${{dD}\over{S(D-1)}}\;1/R$\\
$ $&  $ $ & $ $ \\ \hline
$ $&  $ $ & $ $ \\
$ X^0 \to \infty$ &$\left[{{(D-2) u}\over {2(D-1)}}\right]^{1 \over
{D-2}}~ (X^0)^{2 \over {D-2}}$ &\hspace{4mm} ${{(D-2)u}\over{(D-1)S}}~R $\\
(without string decay) &  $ $ & $ $ \\ \hline
$ $&  $ $ & $ $ \\
$ X^0 \to \infty$ &$\left[{{(D-1) s}\over {2(D-2)}}\right]^{1 \over
{D-1}}~ (X^0)^{2 \over {D-1}}$&usual matter \\
(with string splitting)&  $ $ &  dominated behaviour  \\ \hline
\end{tabular}
\vskip 20pt
\newpage

The unstable string behaviour corresponds to the critical case of the
so-called coasting universe \cite{ell,tur}. In other words, classical
strings provide a {\it concrete} matter realization of such cosmological
model. Till now, no form of matter was known to describe coasting universes
\cite{ell}.

Finally, notice that strings continuously evolve from one type of
behaviour to the other two. This is explicitly seen from the string
solutions in refs.\cite{dms} - \cite{dls} . For example the string
described by $q_-(\sigma, \tau)$ for $ \tau > 0$ shows unstable
behaviour for $ \tau \to 0 $, dual behaviour for $ \tau \to \tau_0 =
1.246450...$ and stable behaviour for $ \tau \to \infty $ .

The equation of state for strings in four dimensional flat Minkowski
spacetime is discussed in ref.\cite{romi}. One finds the values
4/3, 2/3 and 1 for $\gamma$ by choosing  appropriate  values of the
average string velocity in chap. 7 of ref.\cite{romi}.

\section{\bf Self-consistent string cosmology}

In the previous section we investigated the propagation of test
strings in cosmological spacetimes. Let us now investigate how
the Einstein equations in General Relativity and
the effective equations of string theory (beta functions) can be verified
{\bf self-consistently} with our string solutions as sources.

We shall assume a gas of classical strings neglecting interactions as
string splitting and coalescing. We will look for cosmological
solutions described by metrics of the type (\ref{met}). It is natural
to assume that the background will have the same symmetry as the sources.
That is, we assume that the string gas is homogeneous, described by a
density energy $ \rho = \rho(T) $ and a pressure $ p = p(T) $.
In  the  effective equations of string theory we consider a space
independent dilaton field. Antisymmetric tensor fields wil be ignored.

\subsection{\bf String Dominated Universes in General Relativity
(no dilaton field)}

The Einstein equations for the geometry (\ref{met}) take the form

\begin{eqnarray}
{1 \over 2}~(D-1)(D-2)~ H^2 & = & \rho \quad, \nonumber \\
(D-2){\dot H} + p + \rho & = & 0 \quad .
\label{eins}
\end{eqnarray}

where $ H \equiv {{dR}\over{dT}}/ R $. We know $p$ and $\rho$ as
functions of $R$ in asymptotic cases. For large $ R $, the unstable
strings dominate [eq.(\ref{estin})] and we have
\begin{equation}
\rho  = u \; R^{2-D} ~~~,~~~p = -{{\rho} \over {D-1}}\;
\quad {\rm  for ~}R \to \infty
\end{equation}

For small $ R $, the dual regime dominates with

\begin{equation}
\rho  = d \; R^{-D} ~~~,~~~p = +{ {\rho}\over {D-1}}\;\quad {\rm
for ~} R\to 0
\end{equation}

We also know that stable solutions may be present with a contribution
$ \sim R^{1-D} $ to $ \rho $ and with zero pressure. For intermediate values
of $ R $ the form of $ \rho $ is clearly more complicated but a formula
of the type

\begin{equation}
\rho = \left( u_R \; R + {{d} \over R} + s \right) {1 \over
{R^{D-1}}} \label{rogen}
\end{equation}
where
\begin{eqnarray}
\lim_{R\to\infty} u_R = \cases{ 0 \quad & {\rm FRW } \cr
  u_{\infty} \neq 0 & {\rm Inflationary } \cr}
\end{eqnarray}
 This equation of state is qualitatively
correct for all $ R $ and becomes exact for $ R \to 0 $ and $ R \to
\infty $ . The parameters
$u_R , d$ and $ s $ are positive constants and the $u_R$
varies smoothly with $R$.

The pressure associated to the energy density (\ref{rogen}) takes then
the form
\begin{equation}
p  = {1 \over {D-1}} \left( {d \over R} -  u_R \; R\right) {1 \over
{R^{D-1}}} \label{pgen}
\end{equation}

Inserting eq.(\ref{rogen}) into the Einstein-Friedmann equations
[eq.(\ref{eins})] we find
\begin{equation}
{1 \over 2}~(D-1)(D-2)~ \left({{d R}\over{dT}}\right)^2  =  \left( u_R \; R +
 {d \over R} + s \right) {1 \over
{R^{D-3}}} \label{enfri}
\end{equation}

We see that $R$ is a monotonic function of the cosmic time $T$.
Eq.(\ref{enfri}) yields

\begin{equation}
T = \sqrt{{(D-1)(D-2)}\over 2}~
\int_0^R dR \; {{R^{D/2-1}} ~\over {\sqrt{ u_R \; R^2  + d + s\; R}}}
\label{inte}
\end{equation}

where we set $R(0) = 0$.

It is easy to derive the behavior of $R$ for $T \to 0$ and for $T
\to \infty$.

For  $T \to 0$, $ R  \to 0$, the term  $d/R$ dominates in
eq.(\ref{enfri}) and
\begin{equation}
R(T) ~ \buildrel{T \to 0}\over \simeq ~ {D \over 2}
\left[{{2 d}\over {(D-1)(D-2)}}\right]^{1 \over D}~ (T)^{2 \over D}
\label{rcer}
\end{equation}
For $T \to \infty$, $R\to \infty$ and  the term $u_R \,R$ dominates  in
eq.(\ref{enfri}). Hence,
\begin{equation}
R(T) ~ \buildrel{T \to \infty}\over \simeq ~
\left[{{(D-2) u_{\infty}}\over {2(D-1)}}\right]^{1 \over {D-2}}~
(T)^{2 \over {D-2}}
\label{rinf}
\end{equation}
[$u_R$  tends to a constant $u_{\infty}$ for $R\to\infty$].
This expansion is faster than (cold) matter dominated universes where
$ R \simeq  [T]^{{2 \over {D-1}}} $ . For example, for $ D = 4 $, $
R $ grows linearly with $ T $ whereas for matter dominated universes
$ R \simeq  [T]^{2/3} $ .
However,  eq.(\ref{rinf}) {\bf is not} a self-consistent solution.
Assuming that
the term $u_R \,R$ dominates for large $R$ we find a scale factor
$ R(T) \sim  (T)^{2 \over {D-2}} \sim \eta^{2 \over {D-4}}$
for $D\neq 4$ and $ R(T) \sim T \sim e^{\eta}$ at $ D = 4 $.
This {\bf is not an inflationary universe} but a FRW  universe. The
term $u_R  R$ is absent for large $R$ in FRW universes as explained before.
Therefore, we must  instead use for large $R$
\begin{equation}
\rho = \left( {{d} \over R} + s \right) {1 \over
{R^{D-1}}} \label{rocor}
\end{equation}
Now, for  $T \to \infty , R \to \infty $ and we find a matter dominated
regime:
\begin{equation}
R(T)  \buildrel{T \to \infty}\over \simeq
\left[{{(D-1) s}\over {2(D-2)}}\right]^{1 \over {D-1}}~ (T)^{2 \over {D-1}}
\label{mdom}
\end{equation}

For intermediate values of $T~ ,~ R(T)$ is a continuous and
monotonically increasing function of $T$.

In summary, the universe starts at $ T = 0 $ with a singularity of
the type dominated by radiation. (The string behaviour for $R \to 0 $
is like usual radiation).
Then, the universe expands monotonically,
growing for large $ T $ as $ R \simeq  [T]^{{2 \over {D-1}}} $ .
In particular, this gives $ R \simeq  [T]^{2/3} $ for $ D = 4 $.

It must be noticed that the qualitative
form of the solution $R(T)$ does not depend on the particular
positive values of $ u_R , d $ and $ s $.

We want to stress that we achieve a {\bf self-consistent} solution of the
Einstein equations with string sources since the  behaviour
of the string pressure and density given by eqs.(\ref{rogen})-(\ref{pgen})
precisely holds in universes with power like $ R(T) $.

In ref.\cite{cos} similar results were derived using arguments based
on the splitting of long strings.

\subsection{\bf Thermodynamics of strings in cosmological spacetimes}

Let us consider a comoving volume $ R^{D-1} $ filled by a gas of
strings. The entropy change for this system is given by:
\begin{equation}
T dS = d (\rho \; R^{D-1}) ~+ ~ p\;d(R^{D-1})
\label{dentr}
\end{equation}
The continuity equation (\ref{cont}) and (\ref{dentr}) implies that $ dS/dt $
 vanishes.  That is, the entropy per comoving volume stays constant in
time.
 Using now the thermodynamic relation \cite{wei}
\begin{equation}
{{dp}\over{dT}} = {{p+\rho}\over T}
\end{equation}
it follows \cite{romi} that
\begin{equation}
S = {{R^{D-1}}\over T}(p + \rho) + {\rm constant}
\label{entro}
\end{equation}

Let us first ignore the possibility of string decay. Then,
eq.(\ref{entro}) together with  eqs.(\ref{rogen}) and (\ref{pgen}) yields
the temperature as a function of the expansion factor $ R $. That is,
\begin{equation}
T = {1 \over S}\left\{ s + {1 \over {D-1}} \left[ {{D\; d}\over R} +
(D -2)\; u \; R \right] \right\}
\label{tempe}
\end{equation}
where $ S $ stands for the (constant) value of the entropy.

Eq.(\ref{tempe}) shows that for small $ R ,~ T $ scales as  $ 1/R $
whereas for large $ R $ it scales as $ R $. The small $ R $ behaviour
of $ T $ is the usual exhibited by radiation.

For large $ R $, in FRW universes $u_R \to  0 $ and  the constant term
in $s$ dominates. We just find a cold matter behaviour for large $R$.

For large $ R $ in inflationary universes, $u_R \to  u_{\infty}$
 and eq.(\ref{tempe}) would indicate a temperature that {\bf
grows} proportionally to $R$. However, as stressed in ref.\cite{cos},
the decay of long strings  (through splitting) makes $u_R$
exponentially decreasing  with $R$.

\section{\bf Effective String  Equations with the String Sources Included}

Let us consider now the cosmological equations obtained from the low
energy string effective action including the string matter as a
classical source. In D spacetime dimensions, this action can be
written as
\begin{eqnarray}
S & = & S_1 + S_2 \cr
S_1 & = &  {1 \over 2} \int d^Dx \; \sqrt{-G} ~e^{-\Phi}~\left[\; R +
G_{AB}\; \partial^A\Phi~\partial^B\Phi~+ 2~U(G,\Phi) - c\;\right] \cr
S_2 & = &  -{1 \over {4\pi\alpha'}}\sum_{strings}\int d\sigma d\tau~
G_{AB}(X)~\partial_{\mu}X^A\;\partial^{\mu}X^B \qquad,
\label{accef}
\end{eqnarray}
Here $ A, B = 0, \ldots , D-1 $.
This action is written in the so called `Brans-Dicke frame' (BD)
or `string frame', in which matter couples to the metric tensor in the
standard way. The BD frame metric coincides with the sigma model
metric to which test strings are coupled.

Eq.(\ref{accef}) includes the dilaton field ($\Phi$) with a
potential $U(G,\Phi)$ depending on the dilaton and graviton
backgrounds; $ c $ stands for the central charge deficit or
cosmological constant term. The antisymmetric tensor field was not
included, in fact it is irrelevant   for the results obtained here.
Extremizing the action (\ref{accef}) with respect to $G_{AB}$ and $\Phi$
yields the equations of motion
\begin{eqnarray}
R_{AB} + \nabla_{AB}\Phi  + 2 \, {{\partial U}\over {\partial
G_{AB}}} - {{G_{AB}}\over 2} \left[ R + 2 \, \nabla^2\Phi -
(\nabla \Phi)^2 - c + 2 \, U \right] & = & e^{\phi}~T_{AB} \nonumber \\ \cr
R + 2 \, \nabla^2\Phi - (\nabla \Phi)^2 - c + 2 U -
{{\partial U}\over {\partial \Phi}} & = & 0 ~,
\label{eqef}
\end{eqnarray}
which can be more simply combined as
\begin{eqnarray}
R_{AB} + \nabla_{AB}\Phi  + 2 \; {{\partial U}\over {\partial
G_{AB}}} - G_{AB}\;
{\,{\partial U}\over {\partial \Phi}}  & = & e^{\Phi}~T_{AB}
\nonumber \\ \cr
R + 2 \; \nabla^2\Phi - (\nabla \Phi)^2 - c + 2 \, U -
{{\partial U}\over {\partial \Phi}} & = & 0
\label{eqeff}
\end{eqnarray}
Here $ T_{AB} $ stands for the energy momentum tensor of the strings
as defined by eq.(\ref{tens}).
It is also convenient to write these equations as
\begin{equation}
R_{AB}- {{G_{AB}}\over 2}\;R =   T_{AB} + \tau_{AB}
\end{equation}
where $ \tau_{AB} $ is the dilaton  energy momentum tensor :
\begin{eqnarray}
\tau_{AB} & = &  -\nabla_{AB}\Phi + {{G_{AB}}\over 2}
 \left[ 2 \, {{\partial U}\over {\partial \Phi}} - R \right] \nonumber
\end{eqnarray}
The Bianchi identity
\begin{eqnarray}
\nabla^A\left(R_{AB}- {{G_{AB}}\over 2}\;R \right)  & = & 0
\nonumber \cr
\end{eqnarray}
yields, as it must be, the conservation equation,
\begin{equation}
\nabla^A\left( T_{AB} + \tau_{AB}\right) = 0
\end{equation}
It must be noticed that eqs.(\ref{eqeff}) do not reduce to the
Einstein equations of General Relativity even when $ \Phi = U = 0 $.
Eqs. (\ref{eqeff}) yields in that case the
Einstein equations {\it plus} the condition $ R = 0 $.

\subsection{Effective String Equations in Cosmological Universes}

For the homogeneous isotropic spacetime
geometries described by eq.(\ref{met}) we  have
\begin{eqnarray}
R_0^0 & = & - (D-1) ( {\dot H} + H^2 ) \cr
R_i^k & = & - \delta_i^k ~ [ {\dot H} + (D-1)\,H^2 ] \cr
R  & = & - (D-1) ( 2\;{\dot H} + D\,H^2 ) .
\end{eqnarray}
where $H \equiv {1 \over R}\,{{dR}\over{dT}}$ .

The equations of motion (\ref{eqeff}) read
\begin{eqnarray}
{\ddot \Phi } - (D-1) ( {\dot H} + H^2 ) - {{\partial U}\over
{\partial \Phi}} & = &  e^{\Phi}~\rho  \nonumber \\ \cr
 {\dot H} + (D-1)\,H^2 - H \, {\dot \Phi }  + {{\partial U}\over
{\partial \Phi}} + { R \over {D-1}}{{\partial U}\over
{\partial R}} & = &  e^{\Phi}~p  \nonumber \\ \cr
2\;{\ddot \Phi } + 2(D-1)\,H \, {\dot \Phi }- {\dot \Phi }^2
 - (D-1) ( 2\;{\dot H} + D\,H^2 ) - 2\; {{\partial U}\over
{\partial \Phi}} - c + 2\, U  & = & 0
\label{eqcos}
\end{eqnarray}
where dot $ {}^. $ stands for $ {{d}\over{dT}}$, and
\begin{equation}
\rho = T_0^0  \qquad , \qquad -\delta_i^k~p = T_i^k ~.
\end{equation}
The conservation equation takes the form of eq.(\ref{cont})
\begin{equation}
\dot{\rho} + (D-1)\, H\, (p + \rho ) = 0 ~~.
\end{equation}
By defining,
\begin{eqnarray}
\Psi & \equiv & \Phi - \log{\sqrt{-G}} = \Phi - (D-1)\,\log R \cr
{\bar \rho } & =  & e^{\Phi}~\rho \quad, \quad {\bar p } =  e^{\Phi}~p ~,
\label{psiba}
\end{eqnarray}
 equations (\ref{eqcos}) can be expressed in a more compact form as
\begin{eqnarray}
{\ddot \Psi } - (D-1)\,  H^2  - \left.{{\partial U}\over
{\partial \Psi}}\right|_R & = &  {\bar \rho }  \nonumber \\ \cr
 {\dot H}  - H \, {\dot \Psi }  +  \left.{ R \over {D-1}}{{\partial U}\over
{\partial R}}\right|_{\Psi} & = &    {\bar p }  \nonumber \\ \cr
{\dot \Psi }^2 - (D-1)\,H^2  - 2\; {\bar \rho } - 2\, U  + c & = & 0
\; ,
\label{eqcof}
\end{eqnarray}
The conservation equation reads
\begin{equation}
\dot{{\bar\rho}} -  {\dot \Psi }\; {\bar \rho }+ (D-1)\, H\,  {\bar p }
= 0
\end{equation}

As is known, under the duality transformation $ R \longrightarrow
R^{-1} $ , the dilaton transforms as $\Phi \longrightarrow
\Phi + (D-1)\,\log R $. The shifted dilaton $\Psi$ defined by
eq.(\ref{psiba}) is invariant under duality.

The transformation
\begin{equation}
 R' \equiv R^{-1} \quad, ~
\end{equation}
implies
\begin{equation}
\Psi' = \Psi \quad, ~ H' = -H \quad, ~ {\bar p' }=-p \quad, ~
{\bar\rho'} = {\bar\rho}
\end{equation}
provided $ u = d $, that is, a   duality invariant  string source.
This is the duality  invariance transformation of eqs.(\ref{eqcof}).

Solutions to the effective string equations have been extensively
treated in the literature \cite{eqef} and they are not our main
purpose. For the sake of completeness, we briefly analyze the limiting
behaviour of these equations for $ R \to \infty$ and $ R \to 0 $.

It is difficult to make a complete analysis of the  effective string
equations (\ref{eqcof}) since the knowledge about the potential $ U $
is rather incomplete. For weak coupling ($ e^{\Phi} $ small ) the
supersymmetry breaking produces an effective potential that decreases
very fast (as the exponential of an exponential of $\Phi$) for
 $\Phi \to -\infty$.

Let us analyze the asymptotic behavior of  eqs.(\ref{eqcof}) for $ R
\to \infty $ and  $ R \to 0 $ assuming that the potential $ U $
can be ignored.  It is easy to see that a power behaviour Ansatz both
for $ R $ and for $ e^{\Psi} $ as functions of $ T $ is consistent
with these equations. It turns out that the string sources do not
contribute to the leading behaviour here, and we find for  $ R \to 0 $
\begin{eqnarray}
R_{\mp} =&  C_1 \; (T)^{\pm 1/\sqrt{D-1}} ~ \to 0\quad,\cr
 e^{\Psi_{\mp}} =& C_2 \; ( T )^{-1} ~ \to  \left\{
\begin{array}{ll}  \infty \\  0 \end{array} \right.
\label{solef}
\end{eqnarray}
Where $ C_1 $ and $  C_2 $ are constants.
Here the branches $(-)$ and $(+)$ correspond to $ T \to 0 $ and to
 $ T \to \infty $ respectively. In both regimes $ R_{\mp} \to 0 $
and $ e^{\Phi_{\mp}} \to 0$.

The potential $ U(\Phi) $ is hence negligible in these regimes. In
terms of the conformal time $ \eta $ , the behaviours (\ref{solef})
result
\begin{eqnarray}
R_{\mp} =&  C_1' \; (\eta)^{\pm {1 \over {\sqrt{D-1} \mp 1}}} \to 0 \cr
 e^{\Psi_{\mp}} =& C_2' \; (\eta)^{-{{\sqrt{D-1}}\over{\sqrt{D-1} \mp
1}}} \to  \left\{
\begin{array}{ll}  \infty \\  0 \end{array} \right.
\label{sefd}
\end{eqnarray}
Where $ C_1' $ and $  C_2' $ are constants.
The branch $(-)$ would describe an expanding non-inflationary
behaviour near the initial singularity $ T = 0 $ , while the branch
$(+)$ describes a `big crunch' situation and is rather unphysical.

 Similarly,   for   $ R \to \infty $ and
 $ e^{\Phi} \to \infty $, we find
\begin{eqnarray}
R_{\mp} =&  D_1 \; (T)^{\mp 1/\sqrt{D-1}} ~ \to \infty \quad,\cr
 e^{\Psi_{\mp}} =& D_2 \; ( T )^{-1} ~ \to  \left\{
\begin{array}{ll}  \infty \\  0 \end{array} \right.
\label{salef}
\end{eqnarray}
Where $ D_1 $ and $  D_2 $ are constants.
Here again, the branches $(-)$ and $(+)$ correspond to $ T \to 0 $ and to
 $ T \to \infty $ respectively, but now in both regimes
$ R_{\mp} \to \infty$ and $ e^{\Phi_{\mp}} \to \infty $. (In this
limit, one is not guaranteed that $ U $ can be consistently
neglected). In terms of the conformal time, eqs.(\ref{salef}) read
\begin{eqnarray}
R_{\mp} =&  D_1' \; (\eta)^{\mp {1 \over {\sqrt{D-1} \pm 1}}} \to \infty \cr
 e^{\Psi_{\mp}} =& D_2' \; (\eta)^{-{{\sqrt{D-1}}\over{\sqrt{D-1} \pm
1}}} \to  \left\{
\begin{array}{ll}  \infty \\  0 \end{array} \right.
\label{safd}
\end{eqnarray}
The branch $(+)$ describes a noninflationary expanding behaviour for $
T \to \infty $ faster than the standard matter dominated expansion,
while the branch $(-)$ describes a super-inflationary behaviour
$\eta^{-\alpha}$, since $ 0 < \alpha < 1 $, for all D.

The behaviours (\ref{solef}) for  $ R_{\mp} \to 0 $ and (\ref{salef})
for $ R_{\mp} \to \infty$ are related by duality $ R \leftrightarrow 1/R $.

\subsection{\bf String driven inflation?}

Let us consider now the question of whether de Sitter spacetime may be
a self-consistent solution of the effective string equations
(\ref{eqcos})
with the string sources included. The strings in cosmological
universes like de Sitter spacetime have the equation of state
(\ref{rogen})-(\ref{pgen}).
Since $ e^{\Psi} = e^{\Phi} \; R^{1-D} $ :
\begin{eqnarray}
{\bar \rho} & = & e^{\Psi} \left( u \; R + {d \over R} + s \right)
\label{denb} \\ \cr
{\bar p} & = &
{{e^{\Psi}} \over {D-1}} \left( {d \over R} -  u \; R\right)
 \label{prba}
\end{eqnarray}

In the absence of dilaton potential and cosmological constant term,
the string sources do not generate de Sitter spacetime as discussed in
sec. III.A. We see that for $ U = c = 0 $ , and $ R = e^{H T} $ ,
eqs.(\ref{eqcof}) yields to a contradiction
 (unless $ D = 0 $ ) for
the value of $\Psi $ , required to be
 $ -H  T \, + $ constant.

A self-consistent solution describing asymptotically de Sitter
spacetime self-sustained by the string equation of state
(\ref{denb})-(\ref{prba}) is given by
 \begin{eqnarray}
 R & = & e^{H T} ~~, ~~ H = {\rm  constant} > 0 ~~,\cr
 2U-c & = & D\; H^2
=  {\rm  constant} \cr
\Psi_{\pm} & = & \mp H T
\pm i\pi + \log{{(D-1)\,H^2}\over {\rho_{\pm}}} \cr
\rho_+ & \equiv & u \quad , \quad \rho_- \equiv d \quad
\label{anti}
\end{eqnarray}
The branch $\Psi_+$ describes the solution for $ R \to \infty $ ( $ T
\to + \infty  ) $, while the branch  $\Psi_-$ corresponds to  $ R \to 0 $
 ( $ T \to - \infty  ) $. De Sitter spacetime with lorentzian
signature self-sustained by the strings necessarily requires a constant
imaginary piece $ \pm i \pi $ in the dilaton field. This makes $
e^\Psi < 0 $ telling us that the gravitational constant $ G \sim
 e^\Psi < 0 $ here describes antigravity.

Is interesting to notice that in the euclidean signature case, i. e.
(+++\ldots++), the Ansatz ${\dot H} = 0,~ 2U-c=$constant, yields a
constant curvature geometry with a real dilaton, but which is of
Anti-de Sitter type. This solution is obtained from
eqs.(\ref{prba})-(\ref{anti}) through the transformation
\begin{equation}
{\hat X}^0 = i T ~~~,~~~{\hat H} = -i H  ~~~,~~~ X^i = X^i ~~~
,~~~\Psi = \Psi
\label{traf}
\end{equation}
which maps the Lorentzian de Sitter metric into the positive definite
one
\begin{equation}
d{\hat s}^2 = (d{\hat X}^0)^2 + e^{{\hat H}{\hat X}^0}~(d\vec{X})^2.
\label{posi}
\end{equation}
The equations of motion (\ref{eqcof}) within the constant curvature
Ansatz $({\dot{\hat H}} = {\ddot \Psi } = 0 )$
are mapped onto the equations
\begin{eqnarray}
 (D-1)\, {\hat H}^2  - \left.{{\partial U}\over
{\partial \Psi}}\right|_R & = &  {\bar \rho }  \nonumber \\ \cr
  {\hat H} \, {{d \Psi}\over {d {\hat X}^0 }}
+  \left.{ R \over {D-1}}{{\partial U}\over
{\partial R}}\right|_{\Psi} & = &    {\bar p }  \nonumber \\ \cr
-({{d \Psi}\over {d {\hat X}^0 }})^2
+(D-1)\,{\hat H}^2  - 2\; {\bar \rho } - 2\, U  + c & = & 0
\; ,
\label{eqhat}
\end{eqnarray}
with the solution
\begin{eqnarray}
 R & = & e^{{\hat H} {\hat X}^0} ~~, ~~ {\hat H} = {\rm  constant} > 0 ~~,\cr
 c -2 \, U & = & D\; {\hat H}^2
=  {\rm  constant} \cr
\Psi_{\pm} & = & \mp {\hat H } {\hat X}^0
+ \log{{(D-1)\,
{\hat H}^2}\over {\rho_{\pm}}} \cr
\rho_+ & \equiv & u \quad , \quad \rho_- \equiv d \quad
\label{antit}
\end{eqnarray}
Both solutions (\ref{antit}) and (\ref{anti}) are mapped one into
another through the transformation (\ref{traf}).

\bigskip

It could be recalled that in the context of (point particle) field
theory, de Sitter spacetime (as well as anti-de Sitter) emerges as an
exact selfconsistent solution of the semiclassical Einstein equations
with the back reaction included \cite{uno} - \cite{dos}. (Semiclassical in this
context, means that matter fields including the graviton are quantized
to the one-loop level and coupled to the (c-number) gravity background
through the expectation value of the energy-momentum tensor $T_A^B$ . This
expectation value is given by the trace anomaly:
$ <T_A^A> = {\bar \gamma} \; R^2  $). On the other hand,
the $\alpha'$ expansion of the effective string action admits anti-de
Sitter spacetime (but not  de Sitter) as a solution when the quadratic
curvature corrections (in terms of the Gauss-Bonnet term) to the
 Einstein action are included \cite{tres}. It appears that the
corrections to the anti-de Sitter constant curvature are qualitatively
similar in the both cases, with $\alpha'$ playing the r\^ole of the trace
anomaly parameter ${\bar \gamma}$\cite{dos}.

The fact that de Sitter inflation with true gravity $ G \sim
 e^\Psi > 0 $ does not emerge as a solution of the effective string
equations does not mean that string theory excludes inflation. What
means is that the effective string equations are not  enough to get inflation.
The effective string action is a low
energy field theory approximation to string theory containing only the
{\it massless} string modes ({\it massless} background fields).

The vacuum energy scales to start inflation (physical or true vacuum)
are typically of the order of the Planck mass \cite{romi} - \cite{linde}
where the effective string action approximation breaks
down. One must consider the massive string modes (which are absent
from the effective string action) in order to properly get the
cosmological condensate yielding de Sitter inflation. We do not have
at present the solution of such problem.

\bigskip

\bigskip

\begin{centerline}
{TABLE 4. Asymptotic solution of the string effective equations}
(including the dilaton).

\bigskip

\bigskip

{\bf EFFECTIVE STRING EQUATIONS}

{\bf  SOLUTIONS IN COSMOLOGY}

\vskip 20pt
\begin{tabular}{|l|l|l|}\hline
Effective String &  $~~R(X^0)\to 0$  &  $~~R(X^0)\to \infty$\\
\hspace{4mm}equations & \hspace{4mm}behaviour &
\hspace{4mm} behaviour \\ \hline
$ $&  $ $ & $ $ \\
\hspace{4mm}$ X^0 \to 0 $ & $~\sim\,(X^0)^{+ 1/\sqrt{D-1}}$ &
$~\sim\,(X^0)^{-1/\sqrt{D-1}}$\\
$ $&  $ $ & $ $ \\ \hline
$ $&  $ $ & $ $ \\
\hspace{4mm}$ X^0 \to \infty$ & $~\sim\,(X^0)^{-1/\sqrt{D-1}}$ &
$~\sim\,(X^0)^{+ 1/\sqrt{D-1}}$ \\
$ $&  $ $ & $ $ \\ \hline
\end{tabular}
\end{centerline}

\section{Multi-Strings and Soliton Methods in de Sitter Universe}

Among the cosmological backgrounds, de Sitter
spacetime occupies a special
place. This is, in one hand  relevant for inflation and on the
other hand string propagation turns to be specially
interesting  there \cite{dvs87} - \cite{igor}.
String unstability, in the sense that the string proper
length grows indefinitely is particularly present in de Sitter.
The string dynamics in de Sitter universe is described by a
generalized sinh-Gordon model with a potential  unbounded from
below \cite{prd}. The sinh-Gordon function $\alpha(\sigma,\tau)$
having a clear physical meaning : $ H^{-1} e^{\alpha(\sigma,\tau)/2} $
determines the string proper length.
Moreover the classical string equations of motion (plus the
string constraints) turn to be integrable in de Sitter universe
\cite{prd,dms}. More precisely, they are equivalent to a
non-linear sigma model on the grassmannian $SO(D,1)/O(D)$ with periodic
boundary conditions (for closed strings). This sigma model
has an associated linear system \cite{zm1} and using it, one can show the
presence  of an infinite number of conserved quantities \cite{dv78}.
In addition, the string constraints imply
a zero energy-momentum tensor and these constraints
are compatible with the integrability.

The so-called dressing method \cite{zm1} in soliton theory
allows to construct solutions of non-linear classically
integrable models using the associated linear
system. In ref.\cite{cdms}  we systematically construct string solutions
in three dimensional de Sitter spacetime.
We start from a given exactly known solution
of the string equations of motion and constraints in de Sitter \cite{dms}
and then we ``dress'' it.
The string solutions reported there indeed apply to cosmic strings
in de Sitter spacetime as well.

 The invariant interval in $D$-dimensional de Sitter space-time is given by
\begin{equation}
ds^2=dT^{2} ~-~ \exp[2HT]~ \sum_{i=1}^{D-1} (dX^{i})^2 .
\end{equation}
 Here $T$ is the so called cosmic time. In terms of the conformal time
$\eta$,
$$
\eta \equiv-{{\exp[-H T]} \o H} \quad , -\infty < \eta \leq 0 ~ ,
$$
 the line element becomes
$$
ds^2 = {1 \o {H^{2} \eta^{2}}}[  d\eta^2 ~-~ \sum_{i=1}^{D-1}
(dX^{i})^{2}\, ]  ~.
$$
 The de Sitter spacetime can be considered as a $D$-dimensional hyperboloid
embedded in a D+1 dimensional flat Minkowski spacetime with coordinates
$(q^0,...,q^D)$ :
\begin{equation}\label{mehip}
ds^2 = {1\o {H^2} }[ - (dq^0)^2 ~+~ \sum_{i=1}^{D} (dq^i)^2 \, ]
\end{equation}
 where
\begin{eqnarray}
q^0 &=& ~\sinh{H T}~ + {{H^2} \o 2} \exp[H T]~ \sum_{i=1}^{D-1}
(X^i)^2 ~~, \cr \cr
q^1 &=& ~\cosh{H T}~ - {{H^2} \o 2} \exp[H T]~ \sum_{i=1}^{D-1}
(X^i)^2 ~~ , \cr \cr
q^{i+1} &=& ~H \exp[H T]~ X^{i} \quad , \quad  1 \leq i \leq D-1, ~
-\infty < T , ~ X^{i} < + \infty .
\end{eqnarray}
 The complete de Sitter manifold is the hyperboloid
$$
-(q^0)^2 + \sum_{i=1}^{D} (q^i)^{2} = 1.
$$
 The coordinates  $(T, X^{i})$  and  $(\eta , X^{i})$  cover only
the half of the de Sitter manifold $q^0 + q^1 > 0$.

 We will consider a string propagating in this D-dimensional
space-time.The string equations of motion (\ref{conouno}) in the metric
(\ref{mehip}) take the form:
\begin{equation}
\partial_{+-}q + (\partial_{+}q . \partial_{-}q)\;q = 0 \quad
{\rm with} \quad q.q = 1 ,
\label{desmo}
\end{equation}
where . stands for the Lorentzian scalar product $a.b \equiv -a_0b_0 +
\sum_{i=1}^{D}a_ib_i ~ ,~x_\pm \equiv {1\o{2}}(\tau \pm \s)$ and
$\partial_{\pm}q = {{\partial q} \o {\partial x_{\pm}}}$.
The string constraints (\ref{conodos}) become for de Sitter universe
\begin{equation}
T_{\pm\pm}= {\partial q \o {\partial x_\pm}}.
{\partial q \o {\partial x_\pm}} = 0           . \label{bincu}
\end{equation}

 Eqs.(\ref{desmo}) describe a non compact O(D,1) non-linear sigma model in
two dimensions. In addition, the (two dimensional) energy-momentum
tensor is required to vanish by the constraints eqs.(\ref{bincu}) .
This system of non-linear partial differential equations can be
reduced by choosing an appropriate basis for the string coordinates
in the ($D+1$)-dimensional Minkowski space time $(q^0,...,q^D)$
to a noncompact Toda model \cite{prd}.

These equations can be rewritten in the form of a chiral field model
on the Grassmanian
$G_{D}=SO(D,1)/O(D)$. Indeed, any element  ${\bf g}\in G_{D}$ can be
parametrized with a real vector $q\rangle $ of the unit pseudolength
\begin{equation}
{\bf g}=1-2 |q\rangle\langle q|J,\ \\\\\ \  \langle q|J|q\rangle =1.
\label{G}
\end{equation}
In terms of $\bf g$, the string equations (\ref{desmo}) have the following form
\begin{equation}
2\,{\bf g}_{\xi \eta }-{\bf g}_{\xi }\,{\bf g\, g}_{\eta }-
{\bf g}_{\eta }\,{\bf g\, g}_{\xi }=0~,
\label{geq}
\end{equation}
and  the conformal constraints  (\ref{bincu}) become
\begin{equation}
\tr \, {\bf g}_{\xi }^{2}=0,\ \ \ \tr \, {\bf g}_{\eta }^{2}=0~.
\label{gcond}\end{equation}

The fact that ${\bf g}\in G_{D}$ implies that $\bf g $ is a
real matrix with the following properties:

\begin{equation}
{\bf g}=J{\bf g} ^{\mbox{t}}J,\ \ \ {\bf g} ^{2}=I,\ \ \ \
tr \, {\bf g} =D-1 \ , \ \ {\bf g} \in SL(D+1, R).
\label{gcond1}\end{equation}
These conditions are equivalent to the existence of the  representation
(\ref{G}).
Equation (\ref{geq}) is the compatibility condition for
the following overdetermined linear system:
\begin{equation}
\Psi _{\xi}=\frac{U}{1-\lambda }\Psi ,\ \
\Psi _{\eta}=\frac{V}{1+\lambda }\Psi ,
\label{psieq}
\end{equation}
where
\begin{equation}
U={\bf g}_{\xi}\, {\bf g},\ \ \ V={\bf g}_{\eta}\, {\bf g} \ \ .
\label{UV}\end{equation}
Or, in terms of the vector $q\rangle $
$$
U=2\,  q_{\xi }\rangle\langle q \,
J -2 \, q\rangle \langle q_{\xi }\, J , \ \
$$
$$
V=2\,  q_{\eta }\rangle\langle q \, J-2 \,
q\rangle \langle q_{\eta }\, J .
$$
Eq. (\ref{G}) can be easily inverted yielding $q$ in terms of the
matrix $g$:
\begin{equation}
q_0 = \sqrt{{g_{00} -1}\over{2}} \quad , \quad
q_i = \sqrt{{1 - g_{ii} }\over{2}} \; \; 1 \leq i \leq D \;({\rm
no~sum ~ over~}i)
\label{qG}
\end{equation}

The use of overdetermined linear systems to solve non-linear
partial differential equations associated to them goes back to refs.
\cite{ist}. (See refs.\cite{sol} -\cite{lib2} for further references).

In order to fix the freedom in the definition
of $\Psi $ we shall identify
\begin{equation}
\Psi (\lambda =0)=\bf g.
\label{l0g}\end{equation}

This condition  is compatible with the above equations since the matrix
function $\Psi $ at the point $\lambda =0$ satisfies
the same equations as $\bf g$. Thus the problem of
constructing exact solutions of the string equations
is reduced to finding  compatible solutions of the linear equations
(\ref{psieq}) such that  ${\bf g}=\Psi(\lambda =0)$ satisfies
the constraints eqs.(\ref{gcond}) and (\ref{gcond1}).

We concentrate below on the linear system (\ref{psieq}) since this is the
main tool to derive new string solutions in de Sitter spacetime.

In ref.\cite{cdms} the dressing method was applied as follows. We started
from the exact
ring-shaped string solution $ q_{(0)} $ \cite{dms} and we find the
explicit solution $ \Psi^{(0)}(\lambda) $ of the
associated linear system, where $\lambda$ stands for the spectral
parameter.
Then, we propose a new solution $ \Psi(\lambda) $ that differs from
$ \Psi^{(0)}(\lambda) $ by a matrix rational in $\lambda$.
Notice that  $ \Psi(\lambda = 0) $ provides in general a new string solution.

We then show that this rational matrix must have at least {\bf four} poles,
 $ \lambda _{0}, 1/\lambda _{0}, \lambda _{0}^* , 1/\lambda _{0} ^* $,
as a consequence of the symmetries of the problem.
The residues of these poles are shown to be one-dimensional
projectors. We then prove that these projectors are formed by vectors
which can all be expressed in terms of an arbitrary complex constant
vector $ |x_0 \rangle $ and the complex parameter $ \lambda _{0} $.
This result holds for arbitrary starting solutions $ q_{(0)} $.

Since we consider closed strings, we impose a $2\pi$-periodicity on
the string variable $ \sigma $ . This restricts $ \lambda _{0} $ to
take discrete values that we succeed to express in terms
of  Pythagorean numbers.
In summary, our solutions depend on two arbitrary complex numbers
contained in  $ |x_0 \rangle $ and two integers $ n $ and $ m $ . The
counting of degrees of freedom is analogous to $2 + 1$ Minkowski
spacetime except that left and right modes are here mixed up  in a
non-linear and precise way.

The vector $ |x_0 \rangle $ somehow indicates the polarization of the
string. The integers $(n,m)$  determine the string winding. They fix
the way in which the string winds around the origin in the spatial
dimensions (here $S^2$ ). Our starting solution $q_{(0)}(\sigma,
\tau)$ is a stable string
winded $n^2 + m^2$ times  around the origin in de Sitter space.

The matrix multiplications involved in the computation of
the final solution were done with the help of  the computer
program of symbolic calculation ``Mathematica''. The resulting solution
$ q(\sigma, \tau) = (q^0, q^1, q^2, q^3) $ is a complicated
combination of trigonometric functions of $ \sigma $ and hyperbolic
functions of $\tau$. That is, these string solitonic solutions
do not oscillate in time. This is a typical feature of string
unstability \cite{dms} - \cite{gsv} - \cite{agn}. The new feature here is
that strings (even stable solutions) do not oscillate neither
for $ \tau \to 0 $, nor for $\tau \to \pm\infty $.

We plot in figs. 1-7 the solutions for significative values
of  $|x_0 \rangle $ and $(m, n)$ in terms of the comoving coordinates
$( T, X^1, X^2 )$
\begin{equation}
T = \frac{1}{H}\log(q^0 + q^1) ~~,~~~ X^1 =\frac{1}{H}
\frac{q^2}{q^0 + q^1} ~~,~~X^2 =\frac{1}{H}
\frac{q^3}{q^0 + q^1} ~~
\label{cooXT}
\end{equation}
The first feature to point out is that our solitonic solutions
describe {\bf multiple} (here five  or three) strings,
as it can be seen from the fact that for a given time $T$
 we find several different values for $\tau$.
That is, $\tau$ is a {\bf multivalued} function of $T$
for any fixed $\sigma$ (fig.1-2). Each branch of $\tau$
as a function of $T$  corresponds to a different string.
This is a entirely new feature for strings in curved spacetime, with
no analogue in flat spacetime where the time coordinate
can always be chosen proportional to $\tau$.
In flat spacetime, multiple string solutions are described
by multiple world-sheets.
Here, we have a {\bf single} world-sheet describing
several independent and simultaneous strings as a consequence of the
coupling with the spacetime geometry.
Notice that we consider {\it free} strings. (Interactions
among the strings as splitting or merging  are not considered).
Five  is the generic number of strings in our dressed
solutions. The value five can be related to the fact
that we are dressing a one-string solution ($ q_{(0)} $) with {\it four}
poles. Each pole adds here an unstable string.

In order to describe the real physical evolution, we eliminated
numerically $ \tau = \tau (\sigma, T) $ from the solution and expressed
the spatial comoving coordinates $X^1$ and $X^2$ in terms of
$T$ and $\sigma$.

We plot $\tau(\sigma, T)$ as a function of $\sigma$ for different fixed values
of $T$ in fig.3-4. It is a sinusoidal-type function. Besides the
customary closed string period $2\pi$, another period appears
which varies on $\tau$. For small $\tau $ , $\tau = \tau(\sigma, T)$ has
a convoluted shape while for larger $ \tau $ (here $  \tau \leq 5
$), it becomes a regular sinusoid. These behaviours reflect very
clearly in the evolution of the spatial coordinates and shape
of the string.

The evolution of the five (and three) strings
simultaneously described by our solution as a function
of $T$, for positive $T$
 is shown in figs. 5-7. One string is stable (the 5th one). The other four are
unstable. For the stable string, $(X^1, X^2)$ contracts in time
precisely as $ e^{-HT} $, thus keeping the proper amplitude
$(e^{HT} X^1, e^{HT} X^2)$ and proper size constant.
For this stable string $(X^1, X^2) \leq {1 \over H}$.
($1/H$ = the horizon radius).  For the other
(unstable) strings, $(X^1, X^2)$ become very fast constant in time,
the proper size expanding as the universe itself like  $ e^{HT} $ .
For these strings $(X^1, X^2) \geq {1 \over H}$.
These exact solutions display remarkably the asymptotic string
behaviour found in refs.\cite{prd,gsv}.

In terms of the sinh-Gordon description, this means that for the
strings outside the horizon the sinh-Gordon function
$\alpha(\sigma,\tau)$
is the same as the cosmic time $T$ up to
a function of $\sigma$. More precisely,
\begin{equation}
\alpha(\sigma,\tau) \buildrel{T >> {1\over H} }\over =
2 H\, T(\sigma, \tau) + \log\left\{2 H^2 \left[ (A^{1}(\sigma)')^2 +
  (A^{2}(\sigma)')^2 \right] \right\} + O(e^{-2HT}).
\label{Tcosa}
\end{equation}
Here $A^1(\sigma)$ and  $A^2(\sigma)$ are the $X^1$ and $X^2$
coordinates outside the horizon. For $T \to +\infty$ these
strings are at the absolute {\it minimum} $\alpha = + \infty $
of the sinh-Gordon potential with infinite size.
The string inside the horizon (stable string) corresponds to the
 {\it maximum} of the potential, $\alpha = 0$.
$ \alpha = 0$ is the only value in which the string
can stay without being
pushed down by the potential to $\alpha = \pm \infty$ and
this also explains why only one stable
string appears (is not possible to put more than one string at
the maximum of the potential without falling down).
These features are {\it generically} exhibited by our
one-soliton multistring solutions, independently of the
particular initial state of the string
(fixed by $|x^0> $ and $(n,m)$).
For particular values of $|x^0> $, the solution describes
three strings, with symmetric shapes from $ T = 0 $, for instance
like a rosette or a circle with festoons (fig. 5-7).

The string solutions presented here trivially embedd on
$D$-dimensional de Sitter spacetime ($D \geq 3$). It must be noticed
that they exhibit the essential physics of strings in D-dimensional
de Sitter universe. Moreover, the construction method used here
works in any number of dimensions.

\newpage

{\bf Figure Captions}:

\bigskip

{\bf Figure 1:} Plot of the function $H T(\tau)$, for two
values of $\sigma$, for $n = 4, |x^0> = (1+i,.6+.4i,.3+.5i,.77+.79i)$.
The function $\tau(T)$ is multivalued,
revealing the presence of five strings.

\medskip

{\bf Figure 2:} Same as fig.1, for $n = 4, |x^0> = (1,-1,i,1). $
Because of a degeneracy, there are now only three strings.

\medskip

{\bf Figure 3:} $\tau = \tau(\sigma,T)$ for fixed $T$
for $n = 4, |x^0>=(1,-1,i,1)$. Three values of HT are
displayed, corresponding to HT=0 (full line), 1
(dots), and 2 (dashed line). For each HT, three curves are
plotted, which correspond to the three strings. They are ordered with
$\tau$ increasing.

\medskip

{\bf Figure 4:} Same as fig. 3 for
$n = 4, x^0>=(1+i,.6+.4i,.3+.5i,.77+.79i).$
a) The five curves corresponding to the five strings at HT=2.
b) The five curves for three values of HT: HT=0 (full line), 1
(dots), and 2 (dashed line).
\medskip

{\bf Figure 5:} Evolution as a function of cosmic time $HT$ of the
three strings, in the comoving coordinates $(X^1,X^2)$,
for $n = 4, |x^0> = (1,-1,i,1)$.
The comoving size of string (1) stays constant
for $H T <-3$, then decreases
around $H T = 0$, and stays constant again after $HT = 1$.
The invariant size of string (2) is constant for negative $HT$,
then grows as the expansion factor  for $HT > 1$,
and becomes identical to string (1).
The string (3) has a constant comoving size for $ HT < -3$,
then  collapses as $e^{-HT}$ for positive $HT$.

\medskip

{\bf Figure 6:} Evolution of three of the five strings for
$ n = 4, |x^0> = (1+i,.6+.4i,.3+.5i,.77+.79i)$.

\medskip

{\bf Figure 7:} Evolution of the three strings
for the degenerate case $n=6, |x^0> = (1,-1,i,1)$.


\begin{thebibliography}{11}

\bigskip


\bibitem{eri92}H.J. de Vega and N. S\'{a}nchez in ``String Quantum
 Gravity and the Physics at the
              Planck Scale'', Proceedings of the Erice Workshop held in June
              1992. Edited by N. S\'{a}nchez, World Scientific, 1993.
              Pages 73-185, and references given therein.

\bibitem{dvs87}H. J. de Vega and N. S\'anchez, Phys. Lett. {\bf B 197}, 320
(1987).

\bibitem{cos} H. J. de Vega and N. S\'anchez,
Phys. Rev. {\bf D50}, 7202 (1994).

\bibitem{prd} H. J. de Vega and N. S\'anchez,
Phys. Rev. {\bf D47}, 3394 (1993).

\bibitem{dms} H. J. de Vega, A. V. Mikhailov and N. S\'{a}nchez, Teor. Mat.
              Fiz. {\bf 94} (1993) 232.

\bibitem{cdms} F. Combes, H. J. de Vega, A. V. Mikhailov and
N. S\'{a}nchez,

Phys. Rev. {\bf D50}, 2754 (1994).

\bibitem{dls} H. J. de Vega, A. L. Larsen and N. S\'anchez,
Nucl. Phys. {\bf B 427}, 643 (1994).

\bibitem{igor} I. Krichever, Funct. Anal. and Appl. {\bf 28}, 21 (1994),

[Funkts. Anal. Prilozhen.  {\bf 28}, 26 (1994)].

\bibitem{din} H. J. de Vega and I. L. Egusquiza,
Phys. Rev. {\bf D49}, 763 (1994).

\bibitem{ads}  A. L. Larsen and N. S\'anchez,
Phys. Rev. {\bf D50}, 7493 (1994).

\bibitem{hawel} S. Hawking and G. F. R. Ellis, `The large scale
structure of the spacetime',

Cambridge Univ. Press, 1973.

\bibitem{sv} N. S\'anchez and G. Veneziano,
Nucl. Phys. {\bf B333}, 253 (1990).

\bibitem{gsv} M. Gasperini, N. S\'anchez and G. Veneziano,

Int. J. Mod. Phys. {\bf A 6},  3853 (1991) and
Nucl. Phys. {\bf B364}, 365 (1991).

\bibitem{ijm} H. J. de Vega and N. S\'anchez,
Int. J. Mod. Phys. {\bf A 7}, 3043 (1992).

\bibitem{ell} G. F. R. Ellis, Banff Lectures 1990, {\it in} Gravitation,

eds. R. Mann and P. Wesson , World Scientific 1991.

\bibitem{twbk}See for a review, T. W. B. Kibble, Erice Lectures at the
Chalonge School

in Astrofundamental Physics, N. S\'anchez editor, World Scientific, 1992.

\bibitem{vil} A. Vilenkin, Phys. Rev. {\bf D 24},  2082 (1981) and
 Phys. Rep. {\bf 121}, 263 (1985).

N. Turok and P. Bhattacharjee, Phys. Rev. {\bf D 29}, 1557 (1984).

\bibitem{bar} J. D. Barrow, Nucl. Phys. {\bf B 310}, 743 (1988).

\bibitem{tse} R.~Myers,  Phys. Lett.   {\bf B199},  371 (1987).

M.~Mueller,  Nucl. Phys.   {\bf B337},  37 (1990).

See for a review: A.A.~Tseytlin in the Proceedings
of the Erice School

``String Quantum Gravity and Physics at the Planck Energy Scale'',

21-28 June 1992, Edited by N. ~S\'anchez, World Scientific, 1993.

\bibitem{wei} S. Weinberg, `Gravitation and Cosmology', J. Wiley, 1972.

\bibitem{romi} E. W. Kolb and M. S. Turner, `The Early Universe',
Addison-Wesley, 1990.

\bibitem{linde} A. D. Linde, `Particle Physics and Inflationary
Cosmology', Harwood (1990).

\bibitem{pol} J. Polchinsky,  Phys. Lett.   {\bf B209}, 252 (1988).

J. Day and  J. Polchinsky,  Phys. Lett.   {\bf B220}, 387 (1989).

\bibitem{mtw} D. Mitchell, N. Turok, R. Wilkinson and P. Jetzer,
Nucl. Phys. {\bf B 315}, 1 (1989).

 D. Mitchell, B. Sundborg and N. Turok,
Nucl. Phys. {\bf B 335}, 621 (1990).

\bibitem{eqef}  See for example,

I. Antoniadis, C. Bachas, J. Ellis and D. V. Nanopoulos,

 Nucl. Phys. {\bf B 328}, 117 (1989) and  Phys. Lett.  {\bf B 257}, 278 (1991).


B. A. Campbell, A. Linde and K. A. Olive,
 Nucl. Phys. {\bf B 355}, 146 (1991).

 B. A. Campbell, N. Kaloper and K. A. Olive,
 Phys. Lett.   {\bf B 277}, 265 (1992).

A.A.~Tseytlin, Mod. Phys. Lett {\bf A 6}, 1721 (1991).

A.A.~Tseytlin and C. Vafa, Nucl. Phys. {\bf B 372}, 443 (1992).

R. Brustein and P. J. Steinhardt, Phys. Lett. {\bf B 302}, 196 (1993).

M. Gasperini and G. Veneziano, Mod. Phys. Lett. {\bf A 8}, 370 (1993),

Phys. Lett.   {\bf B 277}, 256 (1992)
and Astroparticle Physics {\bf 1}, 317 (1993).

E. Raiten, Nucl. Phys. {\bf B 416}, 881 (1994).

R. Brustein and  G. Veneziano, Phys. Lett. {\bf B 329}, 429 (1994).

V. A. Kosteleck\'y and M. J. Perry,  Nucl. Phys. {\bf B 414}, 174 (1994).

See in addition ref.\cite{tse}.

\bibitem{gsw} See for example: \\
M. Green, J. Schwarz, E. Witten,  `Superstring Theory'.

Cambridge University Press. 1987.

\bibitem{grad}I.S. Gradshteyn and I.M. Ryzhik,
Table of Integrals Series and Products,

(Academic Press, New York, fourth edition, 1965).

\bibitem{ll}See for example,

S. Weinberg, `Gravitation and Cosmology', J. Wiley, 1972.

\bibitem{tur} F. M\"uller-Holstein, Class. Quant. Grav. {\bf 3}, 665 (1986).

K. G. Akdeniz et al. Mod. Phys. Lett. {\bf A 6}, 1543 (1991) and

Phys. Lett. {\bf B 321}, 329 (1994).

\bibitem{fhh} M. V. Fischetti, J. B. Hartle and B. L. Hu, Phys. Rev.
{\bf D 20}, 1757 (1979).

\bibitem{uno} S. Wada and T. Azuma, Phys. Lett. {\bf B 132}, 313 (1983).

V. Sahni and L. A. Kofman, Phys. Lett. {\bf A 117}, 275 (1986).

\bibitem{dos} M. A. Castagnino, J. P. Paz and  N. S\'anchez,
Phys. Lett. {\bf B 193}, 13 (1987).

\bibitem{tres} D. G. Boulware and S. Deser,
Phys. Rev. Lett. {\bf  55}, 2656 (1985).

\bibitem{zm1}V. E. Zakharov and A. V. Mikhailov, JETP, {\bf 75}, 1953 (1978).

\bibitem{dv78} H. J. de Vega,  Phys. Lett. {\bf B 87}, 233 (1979).

\bibitem{ist} C. S. Gardner, J. M. Greene, M. D. Kruskal and R. M. Miura,

Phys. Rev. Lett. {\bf 19}, 1095 (1967).

P. D. Lax, Comm. Pure and Appl. Math. {\bf 21} 467 (1968).

\bibitem{sol} M. J. Ablowitz and H. Segur,
``Solitons and the Inverse scattering
transformation'',

SIAM Philadelphia 1981.

V.E. Zakharov, S.V. Manakov, S.P. Novikov and L.P. Pitaevsky,

``Soliton Theory; The Inverse Method'', Nauka, Moscow, 1980.

\bibitem{lib2} A. C.  Scott, F.  Y. F. Chu and D. W. MacLaughlin,
Proc. IEEE, {\bf 61}, 1443 (1973).

G. L. Lamb, Elements of Soliton Theory, J. Wiley, NY (1980).

\bibitem{agn}H. J. de Vega and N. S\'anchez,
Nucl. Phys. {\bf B309}, 552 and 577 (1988).

C. Loust\'o and N. S\'anchez, Phys. Rev. {\bf D47}, 4498 (1993).

\end{thebibliography}
\end{document}